\documentclass[pre,nofootinbib,showpacs,12pt]{revtex4}
\DeclareRobustCommand{\baselinestretch{1.3}}   

\usepackage{epsfig}

\begin{document}

\title{Adsorption on a Surface with Varying Properties}

\author{A.~S.~Usenko}\email{usenko@bitp.kiev.ua}
\affiliation{Bogolyubov Institute for Theoretical Physics,
Ukrainian National Academy of Sciences, Kiev 03680, Ukraine}


\bigskip

\begin{abstract}

We propose a self-consistent model taking into account variations in adsorption
properties of the adsorbent surface in the process of adsorption--desorption of
molecules of gas on it. We introduce a dimensionless coupling parameter that
characterizes the interaction of an adsorbed molecule with polarized medium. It
is established that the system can be bistable if the coupling parameter is
greater than a critical value and the concentration of gas belongs to a certain
interval. We show that the adsorption isotherms obtained within the framework
of the proposed model essentially differ from the Langmuir isotherms and
establish that the Zeldovich hysteresis is possible. The kinetics of the
surface coverage is analyzed in detail. We show that taking account of
variations in adsorption properties of the surface in the process of
adsorption--desorption leads to new phenomena: a ``quasistationary'' state in
the case of the overdamped approximation and damped self-oscillations of the
system in the general case.

\end{abstract}

\bigskip

\pacs{68.43.-h;  68.43.Mn;  68.43.Nr}

\maketitle

\bigskip



\section{Introduction }  \label{Introduction}

The study of adsorption of molecules on the surface of different bodies covers
an extremely wide class of problems of physics and chemistry and is one of the
most important problems both from the theoretical point of view and for
practical applications. The results of numerous investigations show that
adsorption of molecules on surfaces of bodies leads to changes in various
physical and chemical characteristics of these bodies. The detailed analysis of
changes in the properties of the surface due to adsorption is given in
\cite{ref.Mor,ref.RobMc,ref.Adam,ref.KiK,ref.Vol,ref.Zan,ref.Zhd,ref.LNP,ref.Nau,ref.VaS}.
The results of investigation of changes in properties of the surface due to
adsorption--desorption processes are also widely used in the design of various
sensors (physical, chemical, and biological) \cite{ref.MyaSKZ,ref.Cat} whose
action is based on the use of the change in a certain characteristic of a
sensitive element of the sensor due to adsorption of molecules on its surface.

The results of the theory of adsorption are extremely important for
investigation of heterogenous-catalytic reactions because processes of
adsorption and desorption are integral stages of these reactions.

The classical Langmuir theory that describes adsorption of a gas on solid
surfaces is based on several assumptions. Numerous theoretical investigations,
which, to a large extent, were stimulated by many experimental data that did
not agree with conclusions of the Langmuir theory, were aimed at the
construction of more general models free of one or several restrictions of the
Langmuir theory. An extensive material obtained on the basis of these models
and applications to various problems of adsorption and catalysis are widely
presented in the literature (see, e.g.,
\cite{ref.Zhd,ref.FrK,ref.Bar,ref.Rog,ref.Roz,ref.KrSh,ref.Tov}). In
particular, it is established that taking account of lateral interactions
between adsorbed molecules can lead to a qualitative change in adsorption
isotherms, namely, to a hysteresis of isotherms and to structural changes in
the surface of bodies (surveys of theoretical and experimental results are
given, e.g., in \cite{ref.Adam,ref.Zan,ref.Zhd,ref.LNP,ref.Nau,ref.Tov}).

At the same time, as early as in 1938, in \cite{ref.Zeld}, Zeldovich has
suggested an idea on a change in the surface in the course of adsorption and
desorption due to the presence of adsorbed molecules on it. Using this idea, he
has predicted a hysteresis of adsorption isotherms if the typical time of
adsorption and desorption is much less than the relaxation time of the surface.

Note that a change in adsorption isotherms due to lateral interactions between
adsorbed molecules can also be interpreted as a consequence of a certain change
in adsorption properties of the surface caused by adsorption. However, as far
as we know, the problem of variation in adsorption properties of the surface
itself in the course of adsorption and desorption of molecules of gas on it in
the absence of interactions between adsorbed molecules and the possibility of
hysteresis of adsorption isotherms in the this case remains open.

The present paper is devoted to investigation of specific features of the
behavior of adsorption isotherms and the kinetics of the surface coverage by
adsorbate molecules with regard for variations in adsorption properties of the
surface in the course of adsorption--desorption.

In Sec.~2, we propose a  self-consistent model taking into account variations
in adsorption properties of the surface in the process of
adsorption--desorption of molecules of gas on it. We introduce a dimensionless
coupling parameter that characterizes the interaction of an adsorbed molecule
with polarized medium. We obtain adsorption isotherms and establish that their
behavior essentially depends on the value of this parameter (Sec.~3). It is
shown that, within the framework of the proposed model, the Zeldovich
hysteresis is possible. In Sec.~4, we investigate specific features of the
kinetics of the surface coverage. It is established that variations in
adsorption properties of the surface in the course of adsorption--desorption
cardinally change the Langmuir kinetics.



\section{Model of the Surface with Varying Adsorption Properties }  \label{Modell}

We consider a problem of adsorption of molecules of a one-component gas on the
surface of a solid adsorbent. According to the classical Langmuir theory,
molecules of gas are adsorbed on adsorption centers located on the adsorbent
surface and the number of centers does not change with time. Furthermore, all
centers have equal adsorption activity (energy-uniform surface), do not
interact with each other, and each adsorption center can be bound only with one
adsorbate molecule. The Langmuir kinetics of the quantity of adsorbed substance
is described by the differential equation \cite{ref.FrK}

\begin{equation}
 \frac{d \theta}{dt} = k_a C \bigl( 1 - \theta \bigr) - k_d \, \theta,
\end{equation}

\noindent where $\theta(t) = N_b(t)/N$ is the surface coverage by adsorbate,
$N$ is the total number of adsorption centers, $N_b(t)$ and $N_0(t)$ are,
respectively, the numbers of occupied and free ($N_0(t) = N - N_b(t)$)
adsorption centers at the time  $t$, $k_a$ and $k_d$ are the adsorption and
desorption rate constants, respectively, and $C$ is the concentration of
molecules in the gas phase that is kept constant.

The solution of this equation with zero initial condition $\theta(0) = 0$ has
the form \cite{ref.Roz}

\begin{equation}
 \theta(t) = \theta^{st}_L \biggl[ 1 - \exp\Bigl(-\frac{t}{\tau_{ad}} \Bigr) \biggr],
\end{equation}

\noindent where

\begin{equation}
 \theta^{st}_L = \frac{l}{1 + l}
\end{equation}

\noindent is the stationary surface coverage (Langmuir isotherm), which is
defined by the single dimensionless quantity (dimensionless concentration) $l =
C K$, $K = k_a/k_d$ is the adsorption--desorption equilibrium constant for the
given concentration of gas,

\begin{equation}
 \tau_{ad} = \frac{1}{k_a C + k_d} = \frac{\tau_d}{1 + l}
\end{equation}

\noindent is the time taken for the surface coverage to reach the stationary
value $\theta^{st}_L$, and $\tau_d = 1/k_d$ is the typical lifetime of a
complex (adsorption center + adsorbed molecule).

According to (2) and (3), there is a single-valued correspondence between the
gas concentration and the surface coverage. At the same time, in
\cite{ref.Zeld}, Zeldovich has suggested an idea on a change in adsorption
properties of the surface in the process of adsorption and desorption and,
using this idea, predicted a hysteresis of adsorption isotherms if the typical
time of adsorption and desorption is much less than the relaxation time of the
surface.

In the present paper, within the framework of the Langmuir model, we take into
account variations in properties of the surface of a homogenous adsorbent with
plane boundary in the process of adsorption and desorption of molecules of a
one-component gas on it. We introduce the Cartesian coordinate system centered
at the surface of the adsorbent with $0X$-axis directed into the adsorbent
perpendicularly to its surface so that the adsorbent and the gas occupy the
regions $x \ge 0$ and $x < 0$, respectively. Each adsorption center is
simulated by a one-dimensional linear oscillator of mass $m_0$ that oscillates
perpendicularly to the surface about its equilibrium position $x = 0$.  In the
absence of adsorbate, the motion of an absorption center is described by the
well-known equation of motion of a free linear oscillator

\begin{equation}
 m_0 \frac{d^2x}{dt^2} + \alpha \frac{dx}{dt} + \kappa x = 0,
\end{equation}

\noindent where $\kappa$ is the restoring force constant, $\alpha$ is the
friction coefficient, and $x$ is the coordinate of the oscillator.

In the general case, due to occupation of the adsorption center with a molecule
of adsorbate, the electron structure of the center changes, which results in a
change in the interaction of the center with neighboring atoms of the
adsorbent, i.e., to a local polarization of the adsorbent. As a result, the
medium acts on the complex with a certain force $\vec{F}_p(\vec{r},t)$, where
$\vec{r}$ is the running coordinate of the complex, which is the reaction of
the medium on the electron reconstruction of the adsorption center. Under the
action of this force, the complex tends to a new equilibrium position different
from the equilibrium position  $x = 0$ of the free center. If the polarization
of adsorbent caused by the formation of the complex is axially symmetric about
the axis passing through the complex and parallel to the  $0X$-axis, then this
force has only the component normal to the boundary, $\vec{F}_p(\vec{r},t)=
\vec{e}_x\, F_p(x,t)$, where $\vec{e}_x$ is the unit vector along the
$0X$-axis; for convenience, the center has the coordinate  $\vec{r} = (x, 0,
0)$.

The force  $\vec{F}_p(\vec{r},t)$ either acts on each oscillator of the system
if the oscillator is occupied with a molecule of gas or does not act if it is
free, i.e., the oscillator interacts with the polarized medium only for
discrete time intervals. Instead, we consider an approximation where the
oscillator--medium interaction is continuous in time and the oscillator is
permanently bound with molecule with the time-dependent probability equal to
the surface coverage $\theta$. In this case, $F_p(x,t) = F_p(x)\, \theta$. This
approximation is analogous to the mean-field approximation used in problems of
adsorption with regard for lateral interactions between adsorbed particles
\cite{ref.Zhd}.

We represent the force  $F_p(x)$ in the form $F_p(x) = -
\frac{dU_{int}(x)}{dx}$, where $U_{int}(x)$ is the potential energy of
interaction of the polarized medium with the complex. Expanding the quantity
$U_{int}(x)$ in the Taylor series and keeping only the linear term, we obtain

\begin{equation}
 U_{int}(x) \approx -\chi \, x,
\end{equation}

\noindent where the parameter of complex--polarized medium interaction $\left.
\chi = - \frac{dU_{int}(x)}{dx} \right|_{x = 0}$ is the force acting on the
complex by the medium polarized by this complex.

Ignoring the internal motion in the bound molecule--center system, i.e.,
considering the motion of the complex as a whole, and taking into account a
change in the mass of the oscillator in the process of adsorption--desorption
within the framework of this approximation, we obtain the following equation of
motion for the oscillator:

\begin{equation}
 \frac{d}{dt} \biggl( m_{eff}(\theta) \frac{dx}{dt} \biggr) + \alpha
  \frac{dx}{dt} + \kappa x = \chi \, \theta,
\end{equation}

\noindent where $ m_{eff}(\theta) = m_0 + m \, \theta$ is the effective mass of
the complex that varies in the process of adsorption--desorption and  $m$ is
the mass of an adsorbate molecule. Since $\theta \leq 1$, the effective mass of
the complex does not exceed its total mass $ M = m_0 + m \equiv m_{eff}(1)$.

It follows from Eq.~(7) that bonding of an adsorbate molecule with center leads
to a shift of the equilibrium position of the oscillator by  $(\chi/ \kappa)\,
\theta$ and to a change in the potential energy of the free oscillator equal to
$\kappa x^2/2$ by  $U_{int}(x) \, \theta$. In the limiting case where all
centers are bound, i.e,  $\theta = 1$, the equilibrium position $x_{max} =
\chi/ \kappa$ of the bound oscillator is maximally distant from the surface and
the potential energy of the oscillator $U(x)$ at this equilibrium position is
minimal and equal to a half of the energy of interaction of the bound
oscillator with the polarized medium, $U_{min} \equiv U(x_{max}) = U_{int}/2$,
$\ U_{int} \equiv U_{int}(x_{max})= - \chi^2/\kappa$.

Thus, due to the interaction of adsorbate molecules with adsorption centers,
centers shift relative to the nonperturbed surface of the adsorbent, i.e., this
interaction leads to the formation (for adsorption) and healing  (for
desorption) of local defects of the surface. For $U_{int}(x) <0$, these defects
are ``pits'' (for $\chi > 0$) or ``hills'' (for  $\chi < 0$) whose depth and
height depend on the properties of both the adsorbate and the adsorbent. In the
special case where all atoms of the surface are adsorption centers, this
interaction leads to a shift of the surface of the adsorbent either inwards
(for $\chi > 0$) or outwards (for $\chi < 0$), i.e., to the relaxation of the
surface \cite{ref.Nau}. In other words, the processes of adsorption and
desorption result in a deformation of the surface of the adsorbent, which
leads, in the general case, to changes in the adsorption and desorption rates
and, as a consequence, the surface coverage. Within the framework of the
Langmuir theory of kinetics on the nondeformable surface ($\chi = 0$), the
adsorption and desorption rate constants $k_a$ and $k_d$ do not depend on the
concentration of gas and are defined by the Arrhenius relations

\begin{equation}
 k_a = k_+ \exp{\biggl(-\frac{E_a}{k_B T} \biggr)}, \qquad \qquad
 k_d = k_- \exp{\biggl(-\frac{E_d}{k_B T} \biggr)},
\end{equation}

\noindent where $E_a$ and $E_d$ are the activation energies for adsorption and
desorption, respectively, $k_+$ and  $k_-$ are the preexponential factors, $T$
is the absolute temperature, and $k_B$ is the Boltzmann constant.

A molecule bound with center, due to its interaction with polarized medium, is
in a deeper potential well than in the case of the nondeformable surface.
Therefore, for its desorption, the molecule requires an energy greater than
$E_d$ by the value  $\chi \, x$, where $\chi \, x = |U_{int}(x)|$ is an
additional energy that the bound molecule must acquire to break the bond with
polarized medium.

Generally speaking, the polarization of the medium can also affect the number
of free molecules of gas that can overcome the adsorption barrier $E_a$, i.e.,
a peculiar activation of free molecules of gas occurs and varies in the process
of adsorption--desorption. Here, we do not take into account a change in the
activation energy for adsorption (some results obtained with regard for a
decrease in the activation energy for adsorption in the process of
adsorption--desorption are presented in Appendix A). Supposing that the
preexponential factor $k_-$ is not changed, we obtain the following expression
for  $k_d$:

\begin{equation}
 k_d(x) = k_d \exp{\biggl(-\frac{\chi \, x}{k_B T} \biggr)}.
\end{equation}

\noindent It is worth noting that this quantity already depends on the
concentration of gas because it is defined by the current state of the surface
(the quantity $x$) that depends on the concentration of gas. Therefore,
adsorption and desorption of molecules proceed on the surface whose adsorption
characteristics vary with time.

Introducing the dimensionless coordinate of oscillator  $\xi = x/x_{max}$, we
obtain the following autonomous system of nonlinear differential equations,
which describes the kinetics of the quantity of adsorbed substance with regard
for variations in adsorption properties of the surface in the process of
adsorption--desorption:

\begin{eqnarray}
 \frac{d \theta}{dt} = k_a C \bigl( 1 - \theta \bigr)
  - k_d \, \theta \exp{\left(-g\,\xi \right)}, \\
 \frac{d}{dt} \biggl( m_{eff}(\theta) \frac{d\xi}{dt} \biggr) + \alpha \frac{d\xi}{dt}  =
  \kappa \, \bigl( \theta - \xi \bigr), \
\end{eqnarray}

\noindent where the dimensionless parameter  $g = |U_{int}|/k_B T$, which
characterizes the interaction of an adsorbed molecule with polarized medium,
can be called a coupling parameter. In the absence of interaction (the linear
case) where  $\chi = 0$ ($U_{int} = 0$), the parameter  $g = 0$.

Note that system (10)--(11), in many respects, is analogous to the system of
equations given in \cite{ref.GaKh}, which describes a transport of electrons in
a system of molecules of biological nature with regard for
electron-conformation interaction.



\section{Stationary Case}  \label{Stationary}

In the stationary case, it follows from Eq.~(11) that  $\xi =\theta$.
Therefore, the equilibrium state of system (10)--(11) is defined not by the
pair of quantities  ($\theta^{st}, \xi^{st}$), as is typical of dynamical
systems of two equations \cite{ref.AVKh,ref.BaL}, but only by one quantity
$\theta^{st}$, which is a solution of the equation

\begin{equation}
 l = F(\theta), \qquad \qquad \mbox{where} \qquad\qquad
 F(\theta) = \frac{\theta}{1 - \theta} \, \exp{\Bigl( - g\, \theta \Bigr)}.
\end{equation}

In the general case, it is hardly possible to solve the transcendental equation
(12) in the explicit form. Nevertheless, based on this equation, we can make a
qualitative conclusion on the influence of a change in adsorption properties of
the surface on the surface coverage. To this end, note that the ratio
$\theta/(1 - \theta)$ is equal to  $N_b/N_0$ and the quantity  $l$ is also the
ratio $N_b/N_0$ but in the linear case. Rewriting relation (12) in the form

\begin{equation}
 \frac{N_b}{N_0} = l \, \exp{\Bigl(g\, \theta \Bigr)},
\end{equation}

\noindent we see that a change in properties of the surface caused by
adsorption and desorption leads to an increase in the surface coverage for any
concentration of gas.  The difference between the numbers of bound centers in
the nonlinear ($g \neq 0$) and linear cases increases with the coupling
parameter $g$.

This conclusion can also be made by taking into account that, in the stationary
case, the desorption rate characteristic (9) has the form

\begin{equation}
 k_d(\theta) = k_d \, \exp \bigl(-g\, \theta \bigr),
\end{equation}

\noindent  where the surface coverage $\theta$ is a solution of Eq.~(12). For a
system whose adsorption properties vary in the process of
adsorption--desorption, the adsorption--desorption equilibrium constant

\begin{equation}
 K(\theta) = \frac{k_a}{k_d(\theta)} = K \, \exp \bigl( g\, \theta \bigr)
\end{equation}

\noindent  is greater than the classical adsorption--desorption equilibrium
constant $K $. Therefore, the equilibrium of the system shifts towards an
increase in the number of adsorbed molecules.

It follows from (14) that the interaction of adsorbed molecules with polarized
medium results in an increase in the typical lifetime of complex

\begin{equation}
 \tau_d(\theta) = \frac{1}{k_d(\theta)} = \tau_d \, \exp \bigl(g\, \theta \bigr)
\end{equation}

\noindent and, hence, an increase in the surface coverage.

To analyze solutions of Eq.~(12), we use a standard procedure \cite{ref.BaL},
namely: we consider the plane  $(\theta, l)$ and take into account that  $l
\geq 0$ and $0 \leq \theta \leq 1$ (Fig.~1). The required solutions of Eq.~(12)
are the abscissas of the points of intersection of a horizontal line
corresponding to the given concentration  $l$ (the left-hand side of Eq.~(12))
with the curve $F(\theta)$ (the right-hand side of Eq.~(12)), which is shown in
the figure for different values of the parameter  $g$. Since the behavior of
the function $F(\theta)$ is essentially different for  $g < g_c$ and $g > g_c$,
where $g_c = 4$, it is convenient to represent the parameter $g$ in the form $g
= a_g\, g_c$, where $a_g \geq 0$. For $g < g_c$ (Fig.~1a), the function
$F(\theta)$ monotonically increases and lies to the right of curve 1 for the
linear case ($g = 0$). Thus, as in the linear case, for any given
concentration, the surface coverage has the unique value $\theta^{st}_1 >
\theta^{st}_L$, which agrees with conclusion made above on the basis of
relations (13) and (15). With increase in $a_g$, the curve $F(\theta)$ becomes
more deformed and its deviation from curve 1 increases, which leads to an
increase in the difference $\theta^{st}_1 - \theta^{st}_L$ between the values
of the surface coverage in nonlinear and linear cases. For $a_g = 1$, the
function $F(\theta)$ (curve 1 in Fig.~1b) has an inflection point for  $\theta
= \theta_c = 1/2$ for the concentration $l = l_c = \exp{(-2)} \approx 0.135$.

\begin{figure}
 \centering{
 \epsfig{figure=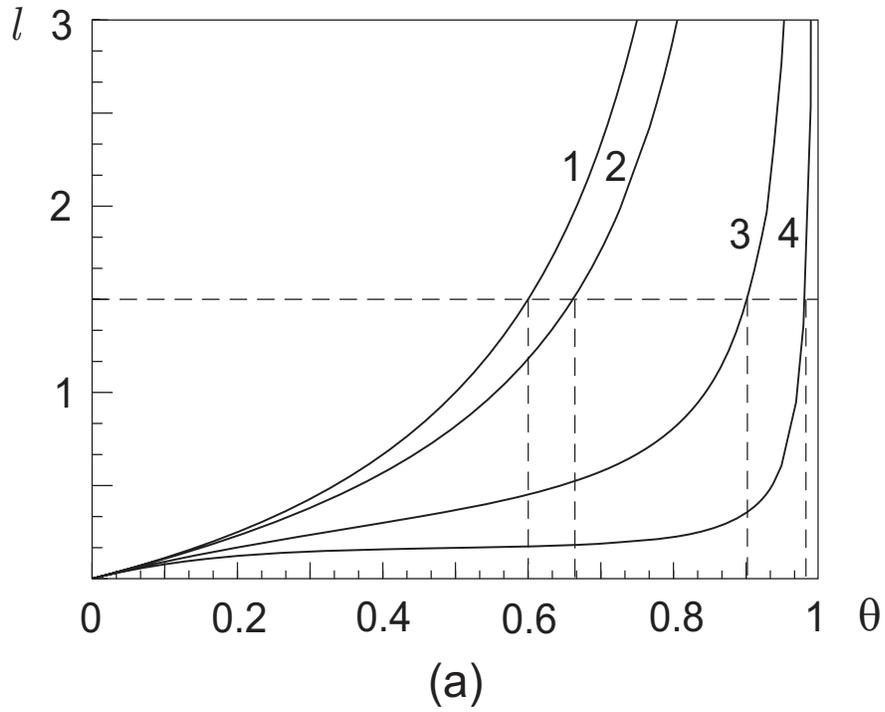, width=12cm, height=9.5cm} \\
  \vspace{0.8cm}
 \epsfig{figure=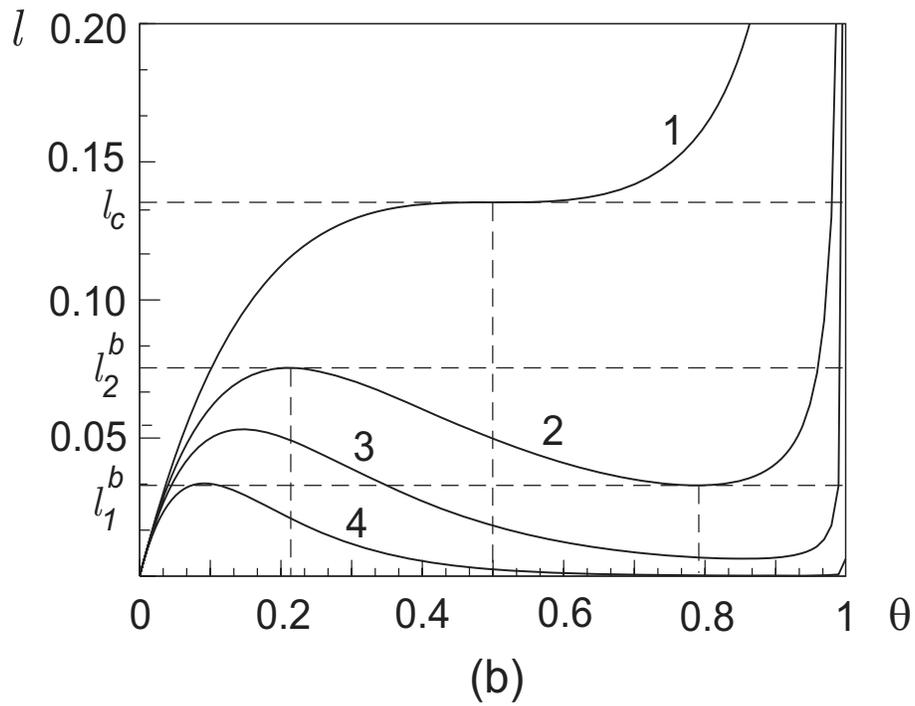, width=12cm, height=9.5cm} \\
  \bigskip
 \caption {Graphical solution of Eq.~(12) for different values of the parameter $a_g$:
  (a) $a_g$ = 0 (1), 0.1 (2), 0.5 (3), 0.9 (4);
  (b) $a_g$ = 1 (1), 1.5 (2), 2 (3), 3 (4).
  Horizontal dashed straight lines correspond to constant values of the dimensionless
  concentration $l$. }
 }
 \label{fig._1}
\end{figure}

For $a_g > 1$, the behavior of the function $F(\theta)$  essentially changes:
for the concentrations $l^b_1$ and $l^b_2$ ($l^b_1< l^b_2 < l_c$) depending on
the value of the parameter $a_g$, the function $F(\theta)$ has a minimum and a
maximum at the points $\theta = \theta^b_1 > \theta_c$ and $\theta = \theta^b_2
< \theta_c$, respectively, which are roots of the quadratic equation

\begin{equation}
  \theta^2 - \theta + \frac{1}{g} = 0
\end{equation}

\noindent and are equal to

$$
 \theta^b_1 = \frac{1}{2} \Biggl( 1 + \sqrt{1 - \frac{4}{g}}\,\, \Biggr),
  \qquad \qquad
 \theta^b_2 = \frac{1}{2} \Biggl( 1 - \sqrt{1 - \frac{4}{g}}\,\, \Biggr).
$$
\noindent The concentrations  $l^b_1$ and $l^b_2$ corresponding to these
surface coverages are defined as follows:

$$
 \l^b_n = \bigl( g \, \theta^b_n - 1 \bigr)\, \exp{(-g \, \theta^b_n)},
  \qquad \qquad  n = 1,2.
$$

In Fig.~1b, the concentrations  $l^b_1$ and $l^b_2$ and the surface coverages
$\theta^b_1$ and $\theta^b_2$ for them are shown with the use of dashed
straight lines for the special case $a_g = 3/2$.  Thus, for concentrations
$l^b_1 < l < l^b_2$, Eq.~(12) has three solutions $\theta^{st}_1 <
\theta^{st}_2 < \theta^{st}_3$, furthermore, only the first solution
$\theta^{st}_1$ lies near the linear $\theta^{st}_L$. With increase in  $a_g$,
the concentrations $l^b_1$ and $l^b_2$ decrease and the difference between
maximum and minimum solutions $\theta^{st}_3 - \theta^{st}_1$ increases.

Analysis of the system of equations (10)--(11) shows that its stationary
solutions $\theta^{st}_1$ and $\theta^{st}_3$ are asymptotically stable and the
solution $\theta^{st}_2$ is unstable.

If the concentration  $l$ tends to the end point of the interval $[l^b_1, \,
l^b_2]$ (to the value $l^b_1$ or $l^b_2$), then the stable  $\theta^{st}_3$ (or
$\theta^{st}_1$) and unstable  $\theta^{st}_2$ solutions approach each other
and, in the limit  $l = l^b_1$ (or $l = l^b_2$), coalesce into one solution
$\theta^b_1$ (or $\theta^b_2$) (in Fig.~1b, for $a_g = 3/2$, these cases are
shown for curve 2.) Therefore, $l = l^b_1$ and $l = l^b_2$ are the bifurcation
concentrations for which the dynamical system (10)--(11) is structurally
unstable \cite{ref.AVKh,ref.BaL} and has the compound (double) equilibrium
states $\theta^b_1$ and $\theta^b_2$. These special cases should be
investigated in their own rights.

\begin{figure}
 \centering{
 \epsfig{figure=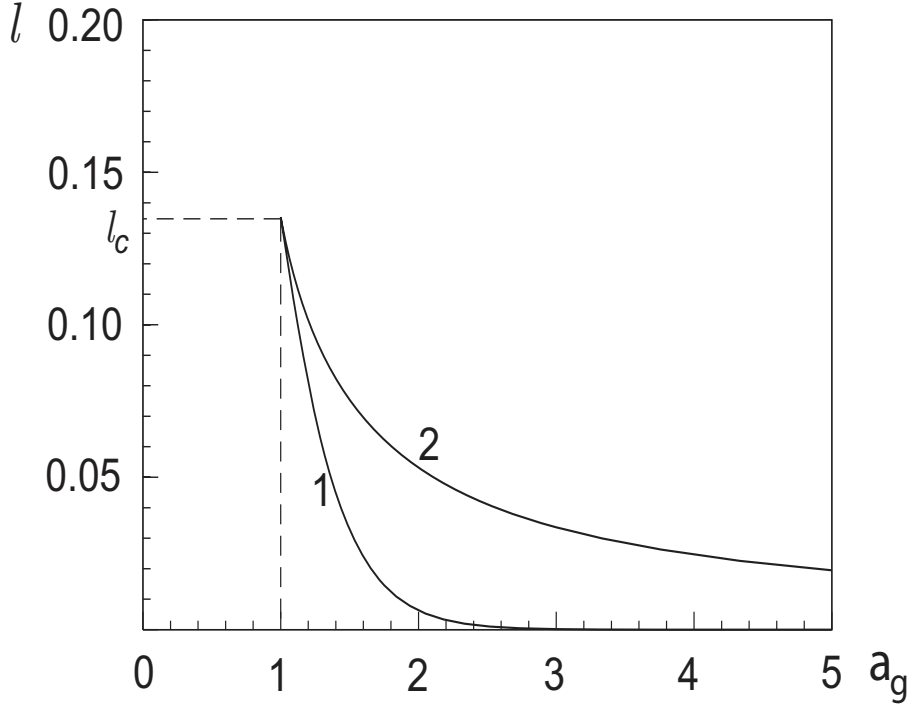, width=12cm, height=9.5cm} \\
  \bigskip
 \caption {Bifurcation curve: branches 1 and 2 of the curve correspond to
  the bifurcation concentrations  $l^b_1$  and  $l^b_2$, respectively. }
 }
 \label{fig._2}
\end{figure}

Using relations (12) and (17), we plot a bifurcation curve in the plane of
parameters ($a_g, \, l$). This curve defined in the parametric form

\begin{equation}
 a_g  =  \frac{1}{4 \, \theta \, \left( 1 - \theta \right) } \, , \qquad\qquad
 l = \frac{\theta}{ 1 - \theta} \, \exp \biggl(- \frac{1}{1 - \theta} \biggr )
\end{equation}

\noindent is shown in Fig.~2.  For any point of this plane lying between the
branches of the bifurcation curve, the system of equations (10)--(11) has three
structurally stable equilibrium states: two states are stable and one is
unstable. If a point lies outside these branches, then the system has one
structurally stable equilibrium state. At any point of the bifurcation curve,
except for the cusp ($a_g = 1$, $l = l_c$), the system has two equilibrium
states: one is structurally stable and another is double \cite{ref.BaL}. At the
cusp, the system of equations (10)--(11) has one triple equilibrium state
\cite{ref.BaL}.

The $S$-shaped adsorption isotherm depicted in Fig.~3 for $a_g > 1$ (curve 1)
essentially differs from the Langmuir isotherm (curve 2) and, on the
qualitative level, reproduces the Zeldovich hysteresis predicted in
\cite{ref.Zeld}.

With quasistatic increase in the concentration from zero, the surface coverage,
at the initial section of the lower stable branch  $0A$ of the isotherm,
coincides with Langmuir one. For these concentrations, a released adsorption
center manages to relax to the nonperturbed state before it binds with other
molecule, furthermore, $\tau_d(\theta) \approx \tau_d$. With increase in the
concentration up to the bifurcation value $l^b_2$, the difference between the
typical lifetimes of complex  $\tau_d(\theta)$ and $\tau_d$ increases. In this
case, a free adsorption center can bind with a subsequent molecule before it
relaxes to the nonperturbed state. In this section of the lower stable branch
of the isotherm, occupation of the surface by adsorbate molecules is determined
by two factors: an increase in the concentration of gas and a change in
adsorption properties of the surface. Due to the last factor, the isotherm
deviates from the Langmuir isotherm, and this deviation increases with
concentration. The pattern cardinally changes as soon as the concentration
negligibly exceeds $l^b_2$. In this case, the lower stable branch of the
isotherm disappears and a new (unique) equilibrium state of the complex is
considerably more distant from the surface than the previous one for $l \le
l^b_2$. Furthermore, the passage to this state is performed for a constant
concentration, i.e., solely due to a change in adsorption properties of the
surface of adsorbent (according to the terminology used in \cite{ref.Zeld}, a
slow adsorption occurs). This passage can require many molecules that
successively take part in the process of adsorption--desorption on the same
adsorption center. Thus, in this stage, a certain interaction between the
molecule leaving the adsorption center and the molecule binding with it occurs.
In Fig.~3, this stage of a sharp increase in the surface coverage for a
constant concentration is shown by the dashed straight line $AB$.

In passing to a stable equilibrium state lying on the upper stable branch of
the isotherm (the point $B$ in Fig.~3), the majority of adsorption centers is
bound. As a result, a subsequent increase in the concentration of gas slightly
affects an increase in the surface coverage. Such a ``saturation'' of the
surface with adsorbate, which rapidly increases with parameter  $a_g$
(Fig.~3b), occurs for concentrations considerably less than those in the linear
case.

\begin{figure}
 \centering{
 \epsfig{figure=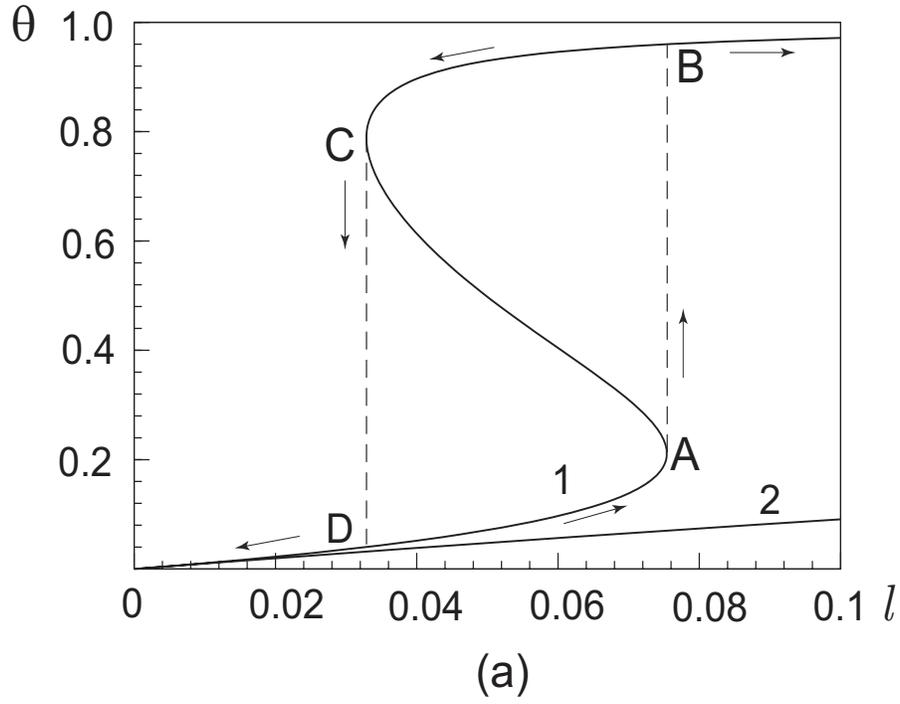, width=12cm, height=9.5cm} \\
  \vspace{1cm}
 \epsfig{figure=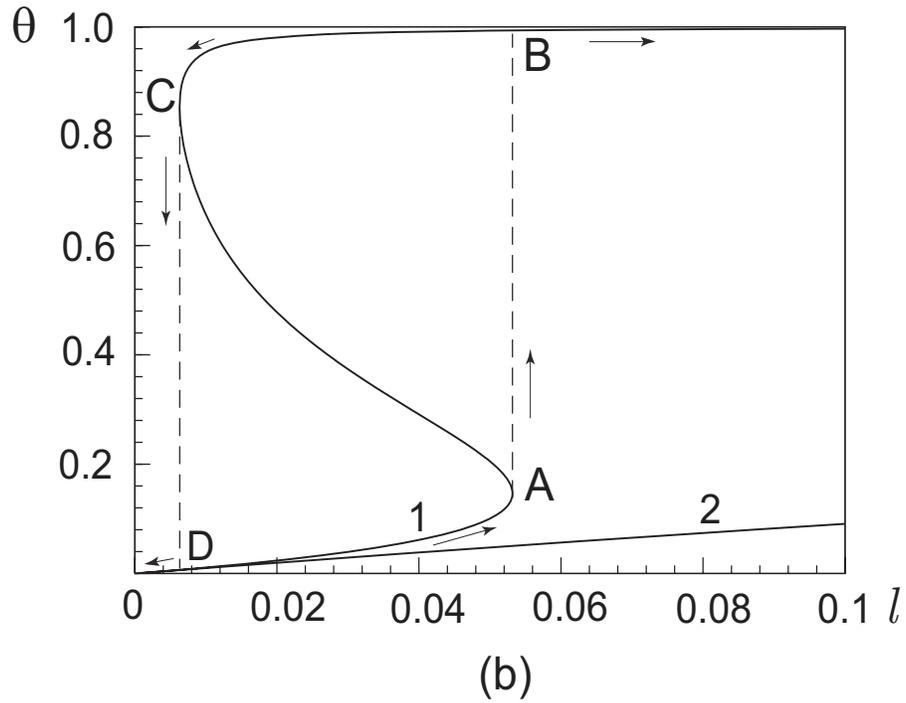, width=12cm, height=9.5cm} \\
  \bigskip
 \caption {Adsorption isotherms for  $a_g$ = 1.5 (a), 2 (b);
  curve 2 stands for the Langmuir isotherm. }
 }
 \label{fig._3}
\end{figure}

In passing through the bifurcation concentration  $l^b_2$, the conditions of
desorption for adsorbed molecules become essentially worse due to a
considerable increase in the depth of the potential well and the displacement
of the equilibrium state of bound adsorption centers from the surface. As a
result, for returning the system to the lower branch of the isotherm, the
concentration should be considerably less than  $l^b_2$. With quasistatic
decrease in the concentration, the surface coverage decreases slightly and
only, in approaching the bifurcation value $l^b_1$, a variation in $\theta$
becomes noticeable. In passing through the bifurcation concentration $l^b_1$,
the upper stable branch of the isotherm disappears and an equilibrium state of
the complex lies considerably closer to the surface than the previous state for
$l \ge l^b_1$. As a result, the surface coverage sharply decreases for the
fixed concentration due to a change (restoration) in properties of the
adsorbent surface. The transition of the system from the upper branch of the
isotherm to its lower branch is shown by the dashed straight line $CD$ in
Fig.~3. Note that this stage of drop of the quantity  $\theta$ is absent in
\cite{ref.Zeld}. A subsequent decrease in the concentration is accompanied by a
decrease in the surface coverage along the lower branch of the isotherm, which,
in fact, coincides with Langmuir isotherm.

This behavior of the adsorption isotherm corresponds to the principle of
perfect delay \cite{ref.Gil} according to which a system, which is in a stable
state at the initial time, with variation in a parameter (concentration in the
case at hand), remains in this state until the state exists.

As the parameter  $a_g$  increases, the bifurcation concentration $l^b_1$
rapidly vanishes (see Fig.~2). Using the results of calculation, we can say
that, for $a_g \geq 3$, a change in adsorption properties of the adsorbent in
the process of adsorption--desorption leads to a peculiar adaptation of the
system to a state in which the majority of adsorption centers are bound up to
very low concentrations.

Note that the isotherms obtained above for the surface whose adsorption
properties vary in the course of adsorption--desorption (Fig.~3) are similar to
the isotherms obtained with regard for lateral interactions between adsorbed
molecules on a nondeformable surface \cite{ref.Adam,ref.Zhd,ref.Tov} and to the
Hill--de~Boer isotherms derived on the basis of the Hill--de~Boer equation of
state for adsorbed molecules (a two-dimensional analog of the Van der Waals
equation) \cite{ref.JaP}.

In analysis of adsorption isotherms with regard for lateral interactions
between adsorbed molecules (see, e.g., \cite{ref.BrMed}), for investigation of
possible surface phase transitions, a critical temperature $T_c$ is introduced
\cite{ref.Zhd,ref.LNP,ref.Nau,ref.JaP}. For the model considered in the present
paper, using the expression for the coupling parameter $g$, the critical value
$g_c$, and the analysis of adsorption isotherms performed above, the critical
temperature is defined as follows: $k_B T_c = |U_{int}|/4$. For a system of
adsorbed molecules, one stable state occurs for  $T > T_c$, whereas, for  $T <
T_c$, two stable states are possible. The corresponding phase diagram for the
adsorbed layer in the ``surface coverage--critical temperature'' coordinates is
determined by the relation

$$
 \frac{T}{T_c}  =  4\, \theta \, \left( 1 - \theta \right) \,
$$

\noindent and, as in the case taking into account lateral interactions between
adsorbed molecules on a nondeformable surface within the framework of the
mean-field approximation  \cite{ref.Zhd}, is symmetric about $\theta = 1/2$.



\section{Nonstationary Case}  \label{Nonstationary}

\subsection{Overdamped Approximation}  \label{Massless}

First, we investigate the kinetics of system (10)--(11) within the framework of
overdamped approximation where the masses of adsorption center and molecule are
low and the friction coefficient is so large that the first term on the
left-hand side of Eq.~(11) can be neglected as compared with the second term.
Using the well-known results for a linear free oscillator of constant mass
\cite{ref.AVKh}, this approximation is correct if

\begin{equation}
 \tau^2_M  \ll \tau^2_r,
\end{equation}

\noindent where  $\tau_M = 1/\omega_M$, $\omega_M = \sqrt{\kappa/M}$ is the
oscillation frequency of an oscillator of mass $M$, and $\tau_r =
\alpha/\kappa$ is the typical relaxation time of a massless oscillator. Since
$M$ is the maximally possible effective mass, condition (19) is even somewhat
high. In this approximation, the system of equations (10)--(11) is simplified
to the form

\begin{eqnarray}
 \frac{d \theta}{dt'} = l \, \bigl( 1 - \theta \bigr)
  - \theta \exp{\left(-g\,\xi \right)}, \\
 \frac{d\xi}{dt'}  = \frac{\theta - \xi}{\beta}, \qquad \qquad \qquad \qquad
  \end{eqnarray}

\noindent where $t' = t/\tau_d$ is the dimensionless time and  $\beta =
\tau_r/\tau_d\,$.

Analysis of system (20)--(21) performed on the basis of the qualitative theory
of differential equations \cite{ref.BaL,ref.ArP} shows that the stable
equilibrium states of the system $\theta^{st}_1$ and $\theta^{st}_3$ are stable
nodes and its unstable equilibrium state $\theta^{st}_2$ is a saddle. For the
bifurcation concentration $l = l^b_1$ (or $l = l^b_2$), the system is
structurally unstable and has a compound equilibrium state $\theta^b_1$ (or
$\theta^b_2$), namely, a saddle-node with two saddle sectors and one stable
nodal sector. The system is also structurally unstable for the critical
concentration $l = l_c$ and $a_g = 1$. In this case, the system has one
equilibrium state $\theta_c$, which is stable triple node.

The numerical analysis of the system of equations with initial conditions for
$t' = 0$

\begin{equation}
 \theta(0) = 0   , \qquad\qquad  \xi(0) = 0
\end{equation}

\noindent shows that, for any values of the parameters  $a_g$, $l$, and
$\beta$, the system monotonically evolutes to the nearest stable equilibrium
state. Therefore, for the bistable system ($a_g > 1$ and $l^b_1 \leq l \leq
l^b_2$), the stable equilibrium state $\theta^{st}_3$ is inaccessible. The time
taken for attaining the equilibrium state  $\theta^{st}_1$ considerably depends
on parameters, first of all, on the concentration.

Let us investigate the kinetics of the surface coverage for a system that can
be bistable for  $a_g = 1.5$. In this case,  $l^b_1 \approx 0.0329$ and $l^b_2
\approx 0.0754$; $\theta^b_1 \approx 0.789$ and $\theta^b_2 \approx 0.211$.

In Fig.~4, the kinetics of the surface coverage $\theta(t')$ is shown for
concentrations less ($l = 0.05$, Fig.~4a) and greater ($l = 0.1$, Fig.~4b) than
the bifurcation concentration $l^b_2$.  For comparison, the Langmuir kinetics
is shown in this figure by curve 2. For $l < l^b_2$, the behavior of
$\theta(t')$ is analogous to that in the Langmuir case: the quantity
$\theta(t')$ monotonically increases from zero to the nearest stationary value
$\theta^{st}_1$ that lies near the linear value  $\theta^{st}_L$ (Fig.~4a).
With increase in the concentration, this behavior remains true up to the
bifurcation value $l^b_2$ (moreover, both the stationary value $\theta^{st}_1$
and the time taken for its attaining increase).

\begin{figure}
 \centering{
 \epsfig{figure=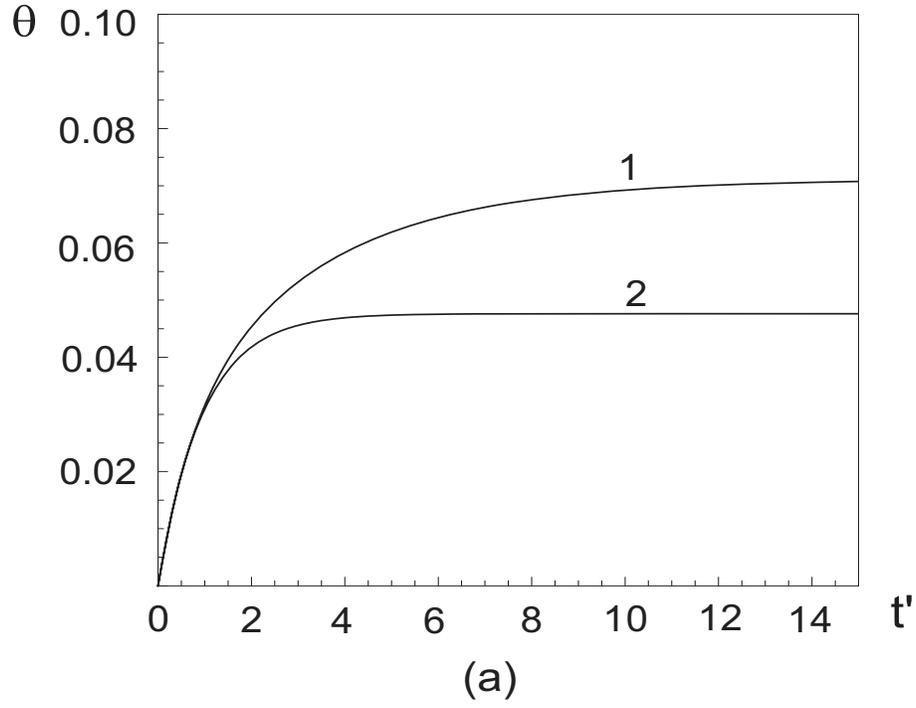, width=12cm, height=9.5cm} \\
  \vspace{1cm}
 \epsfig{figure=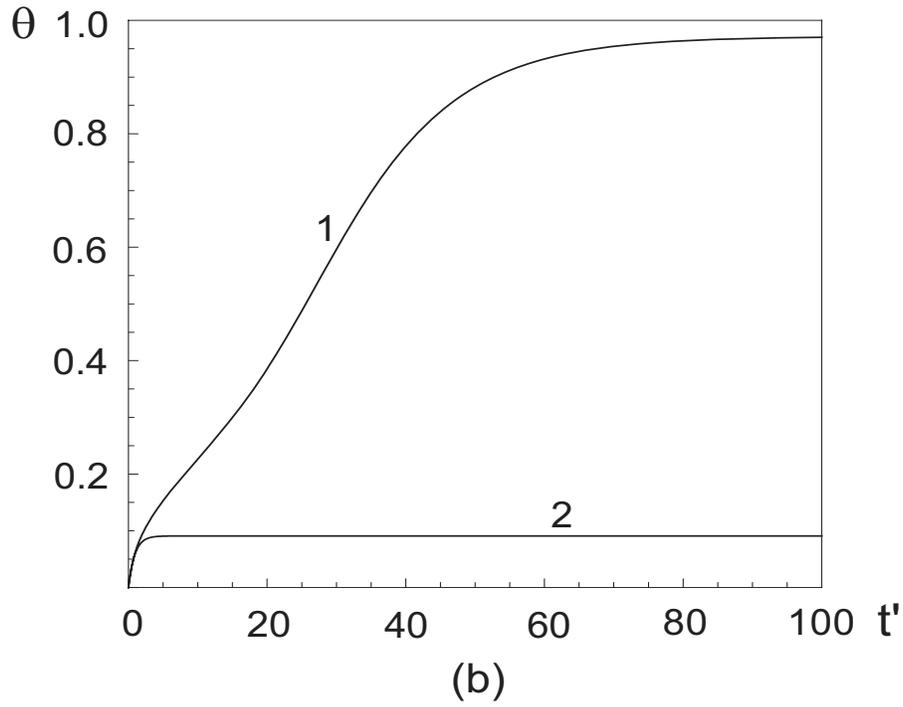, width=12cm, height=9.5cm} \\
  \bigskip
 \caption {Kinetics of the surface coverage  $\theta(t')$ on deformable
  (curve 1) and nondeformable (the Langmuir kinetics, curve 2) surfaces for
  $a_g = 1.5$, $\beta = 1$; $l = 0.05$ (a), $0.1$ (b). }
 }
 \label{fig._4}
\end{figure}

For $l > l^b_2$, the system has only one stable equilibrium state, furthermore,
in this state, the surface coverage, which is close to the maximum possible
value, is essentially greater than that in the linear case.  Moreover, both the
shape of the kinetic curve $\theta(t')$ and the time taken for attaining the
stationary value considerably differ from the Langmuir ones (Fig.~4b). With
increase in the concentration, the time taken for attaining the stationary
value decreases.

\begin{figure}
 \centering{
 \epsfig{figure=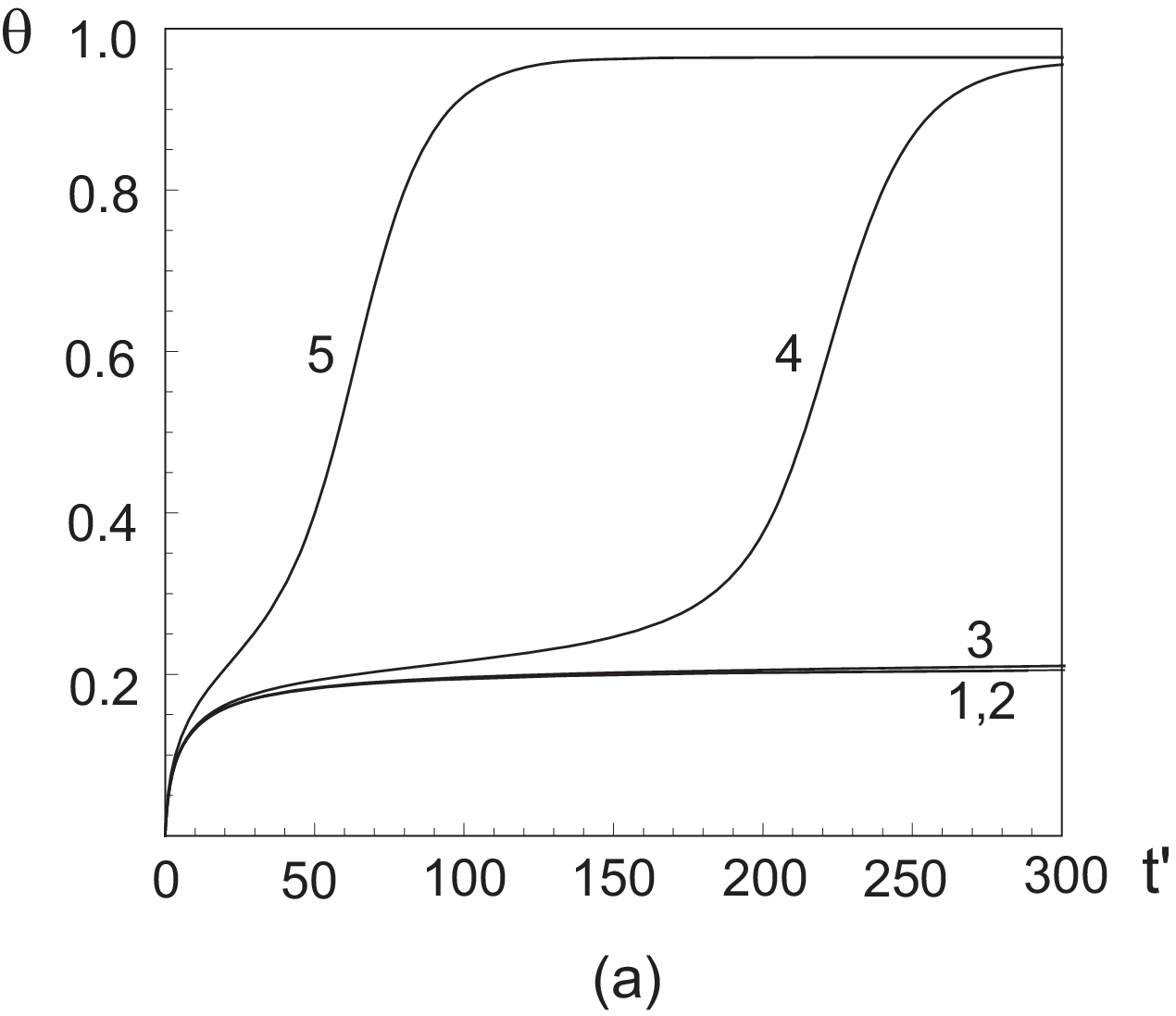, width=12cm, height=9.5cm} \\
  \vspace{1cm}
 \epsfig{figure=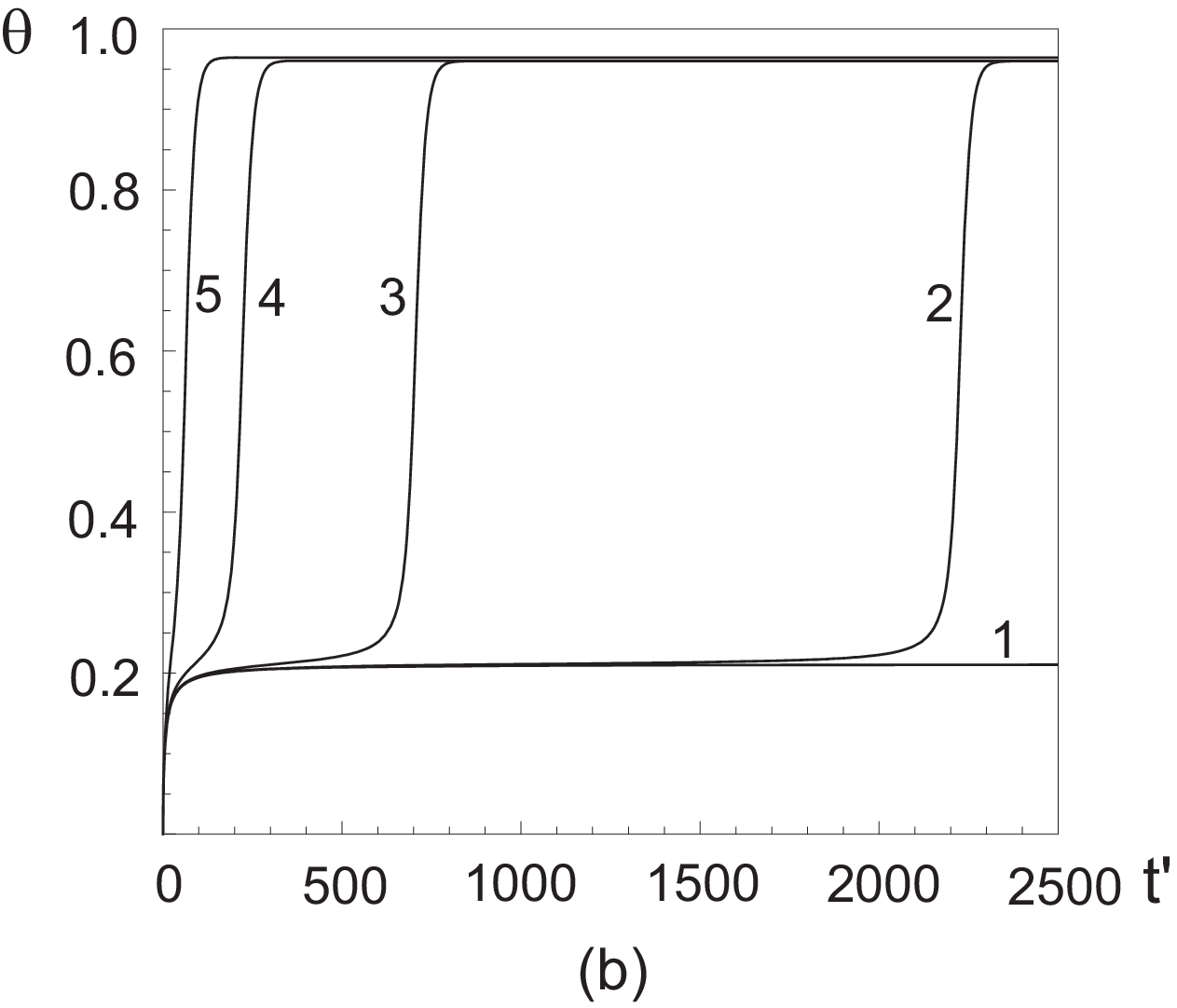, width=12cm, height=9.5cm} \\
  \bigskip
 \caption {Kinetics of the surface coverage  $\theta(t')$ for concentrations
  close to the bifurcation concentration $l^b_2$:
  $a_g = 1.5$, $\beta = 1$;
  $\delta = 0$ (1), $10^{-4}$ (2), $10^{-3}$ (3), $10^{-2}$ (4), $10^{-1}$ (5). }
 }
 \label{fig._5}
\end{figure}

The value of the parameter  $\beta$ affects only the time taken for the system
to attain the stationary value but does not qualitatively change the kinetics
of   $\theta(t')$ both for  $l < l^b_2$ and for $l > l^b_2$. This time
decreases if $\beta$ decreases and increases if $\beta$ increases, which is
quite natural because a variation in $\beta$ is equivalent to a variation in
the friction coefficient $\alpha$.

For concentrations  $l > l^b_2$ near the bifurcation concentration $l^b_2$, the
behavior of  $\theta(t')$ qualitatively changes. The behavior of the quantity
$\theta(t')$  for $l = l^b_2 \, \left(1 + \delta \right)$, for low values of
the relative concentration  $\delta = \left(l - l^b_2\right)/l^b_2 \geq 0$, is
shown in Fig.~5. If the concentration slightly exceeds the bifurcation
concentration $l^b_2$ (curves 2 and 3), then the evolution of  $\theta(t')$ can
be conditionally divided into three stages: (i) from the initial zero value to
a value of  $\sim\theta^b_2$ corresponding to the bifurcation concentration
$l^b_2$; (ii) a very slow (in the limit $\lim\limits_{\delta \rightarrow 0}\,$,
infinitely slow) variation in the neighborhood of $\theta^b_2$; (iii) from
$\sim\theta^b_2$ to the stationary value $\theta^{st}_1$. For low values of
$\delta$, the time taken for attaining the stationary level $\theta^{st}_1$ is
determined mainly by the second (``quasistationary'') stage in which the
system, in fact, does not change (curve 2 in Fig.~5b). This behavior of the
surface coverage $\theta(t')$ is caused by the well-known effect of slowing
down of a system near a singular point for the bifurcation value of a parameter
\cite{ref.Gil,ref.Hak,ref.Sug,ref.Byk} in the case where a phase trajectory of
the system moves near this point.

This behavior of system (20)--(21) can be clearly explained on the basis of
analysis of its phase trajectories in the phase plane  $(\theta, \, \xi)$. The
phase trajectories of the system with zero initial condition (22) are shown for
concentrations less (Fig.~6a), equal (Fig.~6b), and slightly greater (Fig.~6c)
than the bifurcation concentration  $l^b_2$. The dashed lines in these figures
stand for the main isoclines of the system: the isocline of horizontal slopes
$\xi = \theta$ and the isocline of vertical slopes $\xi = (1/g)\, \ln(\theta/l
\, (1 - \theta))$. The points of intersection of these isoclines are singular
points of the system.

\begin{figure}
 \centering{
 \epsfig{figure=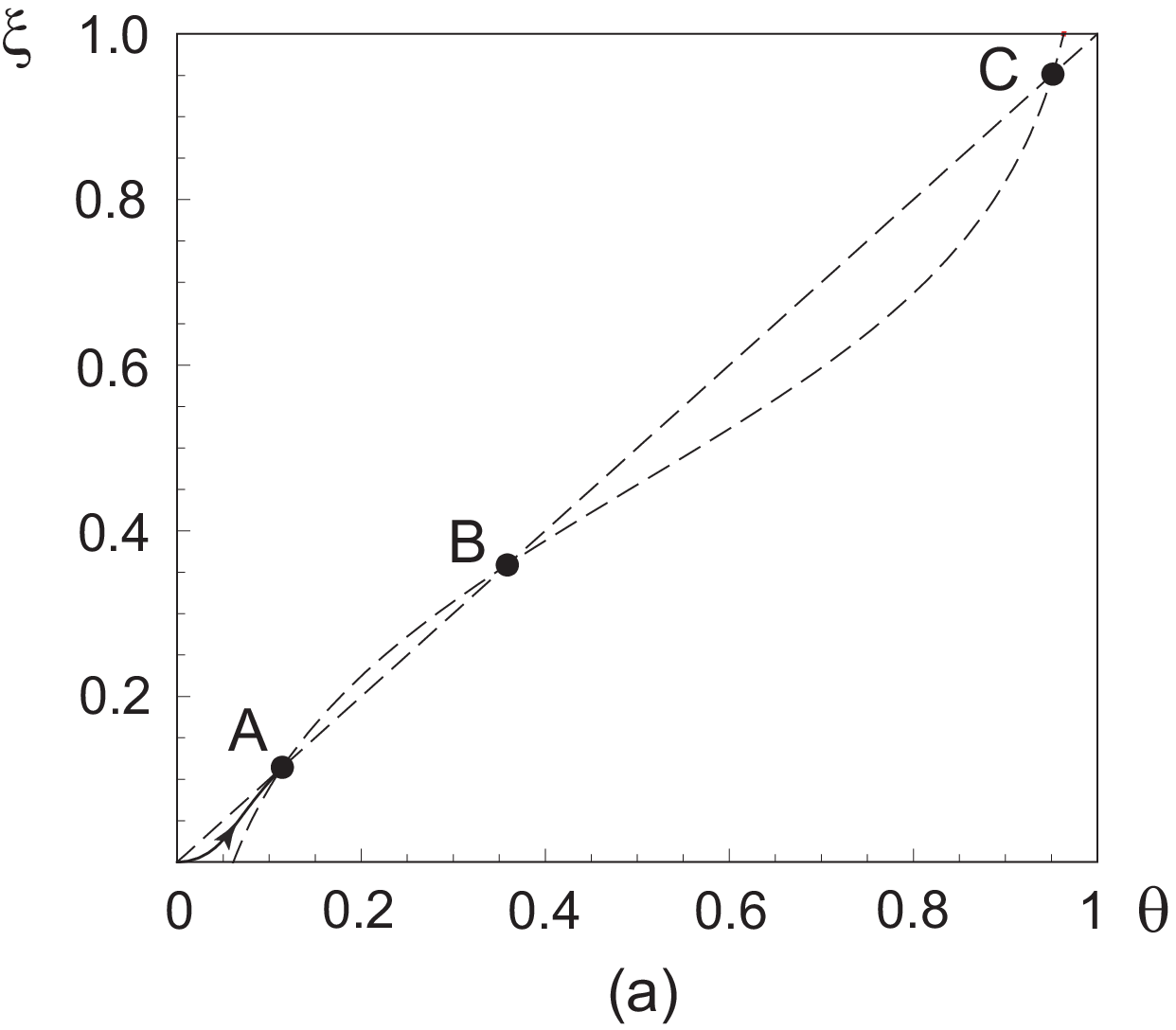, width=7.5cm, height=7.5cm} \hspace{0.5cm}
 \epsfig{figure=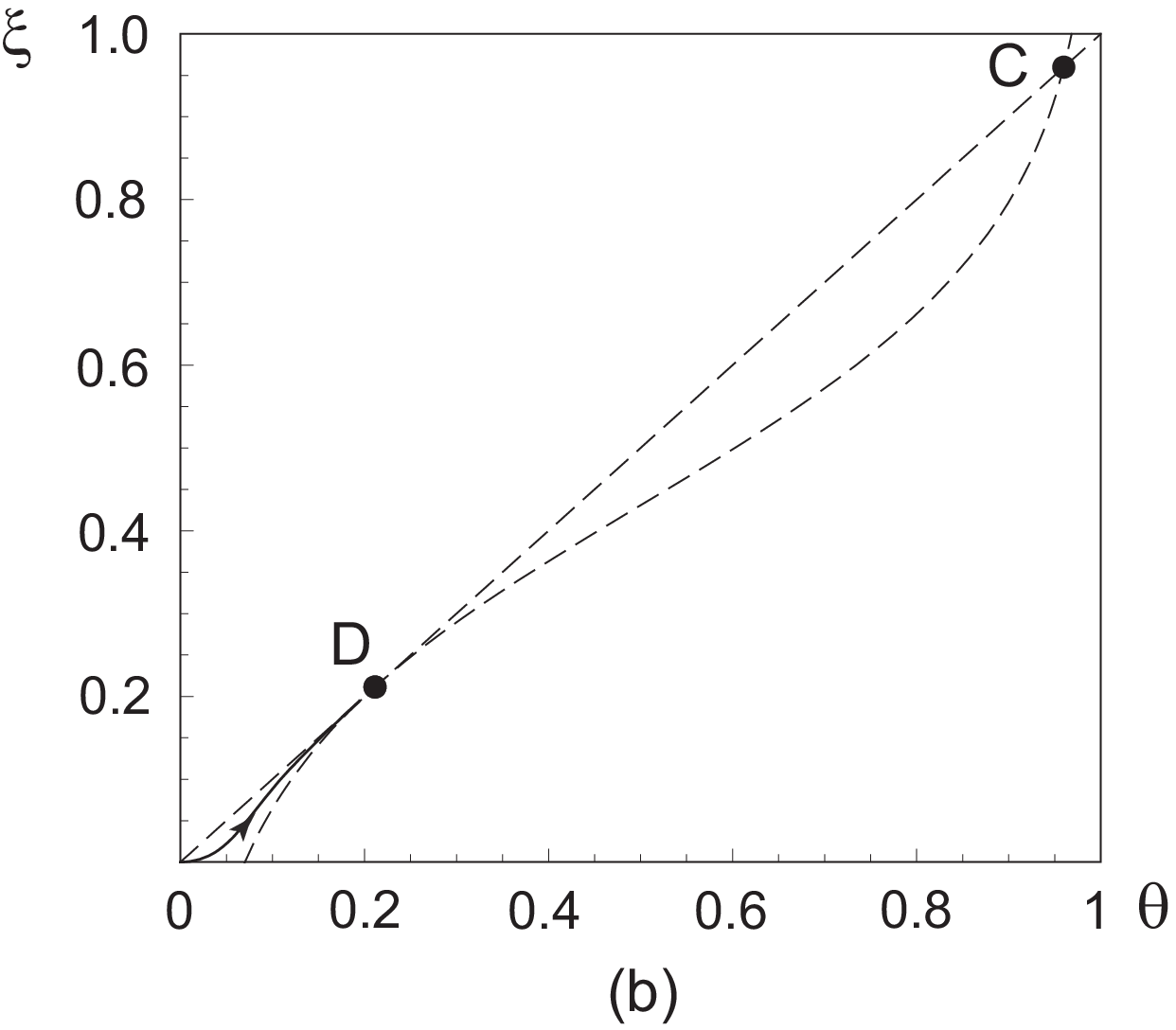, width=7.5cm, height=7.5cm} \\
  \vspace{0.6cm}
 \epsfig{figure=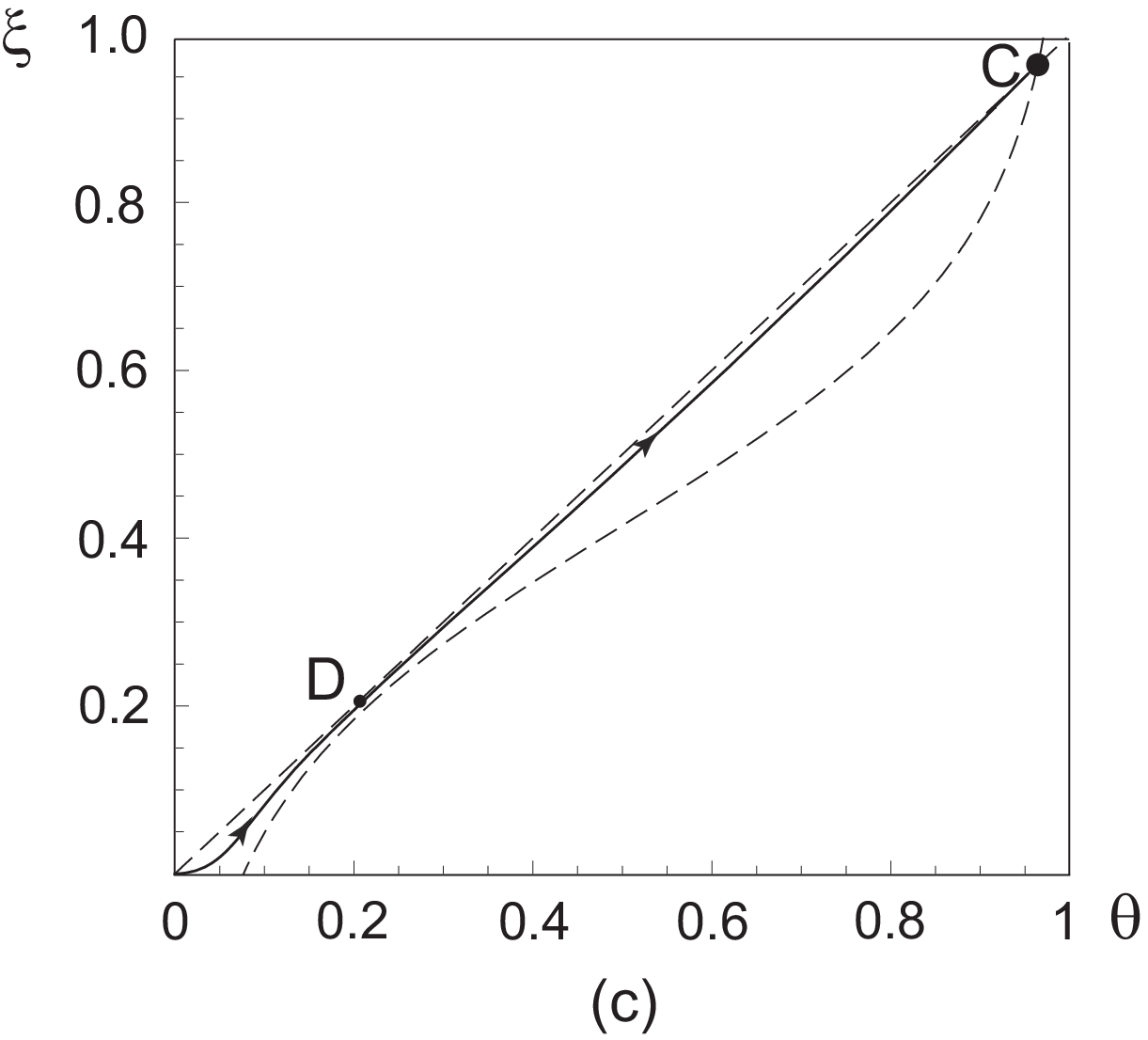, width=7.5cm, height=7.5cm} \\
  \bigskip
 \caption {Phase trajectories of system (20)--(22) for concentrations less ($l = 0.065$) (a),
  equal($l = l^b_2$) (b), and slightly greater
  ($l = l^b_2 \,\left(1 + \delta \right)$,  $\delta = 0.1$) (c)
  than the bifurcation concentration $l^b_2$;
  $a_g = 1.5$, $\beta = 1$.
  Dashed lines stand for the main isoclines of the system.}
 }
 \label{fig._6}
\end{figure}

For $l < l^b_2$ (Fig.~6a), the singular points  $A$ and  $C$ are stable (stable
nodes) and the singular point  $B$ is unstable (saddle). The phase trajectory
in Fig.~6a starting from the origin of coordinates and going to the nearest
singular point $A$ completely lies between the main isoclines. Moreover, the
immediate analysis of system (20)--(21) shows that all phase trajectories of
system (20)--(21) with initial values belonging to the domain bounded by the
sections of the main isoclines before their intersection at the point  $A$ also
completely lie between the main isoclines. A change in the parameter  $\beta$
does not qualitatively change the behavior of the phase trajectories and only
shifts them to one of the main isoclines: for  $\beta << 1$ and $\beta >> 1$,
the phase trajectories are closely pressed to the isoclines of horizontal and
vertical slopes, respectively.

For $l = l^b_2$, the singular points  $A$  and $B$ coalesce into one (compound)
singular point  $D \equiv D(\theta^b_2, \, \xi^b_2)$, which is a point of
tangency of the main isoclines (Fig.~6b). In this case, the phase trajectory is
analogous to that in the previous case. For $l > l^b_2$, the system has only
one (stable) singular point  $C$ (Fig.~6c). If the relative concentration is
low, then a gap between the main isoclines in the neighborhood of their point
of tangency  $D$ for  $l = l^b_2$ is also small. Since a phase trajectory does
not leave the domain bounded by the main isoclines, it goes through the gap
and, in a neighborhood of the point  $D$,  its motion becomes slower.
Furthermore, the less the relative concentration, the narrower the gap between
the main isoclines and the closer the phase trajectory approaches the point $D$
and, hence, the more its slowing down near the point. This behavior of the
system corresponds to the effect of critical slowing down near a degenerate
critical point \cite{ref.Gil,ref.Hak,ref.Sug}.

As a result, for low values of  $\delta$, the function  $\theta(t')$ in Fig.~5b
has the form of a double step (curves 2 and 3). The first plateau of the step
corresponds to the quasistationary state  $\theta^b_2$ and the second
corresponds to the stable state $\theta^{st}_1$.

For $\beta << 1$, the kinetics of the surface coverage, which is shown in
Figs.~4 and 5 for the intermediate case  $\beta = 1$, can be analyzed in a
standard way with the use of a potential  \cite{ref.Gil,ref.Hak}. It follows
from Eq.~(21) that $\xi =\theta$. Substituting this relation into (20), we
obtain the following equation for  $\theta$:

\begin{equation}
 \frac{d \theta}{dt'} = - \frac{dV(\theta; l, g)}{d \theta},
\end{equation}

\noindent where the potential $V(\theta; l, g)$ can be represented as the sum
\begin{equation}
 V(\theta; l, g) = V_L(\theta; l) + V_{ind}(\theta; g),
\end{equation}

\noindent  where

\begin{equation}
 V_L(\theta; l) = \frac{l + 1}{2}\, \theta \left(\theta - 2 \theta_L^{st}\right)
\end{equation}

\noindent is the parabolic potential for the Langmuir kinetics (1) and

\begin{equation}
 V_{ind}(\theta; g) = \frac{1}{g^2}\, \Bigl\{1 - (1 + g\,\theta)\,
  \exp{\left(-g\, \theta \right)} \Bigr\} - \frac {\theta^2}{2}\, .
\end{equation}

\noindent is the potential caused by the action of the polarized medium on the
complex.

Thus, the behavior of the quantity   $\theta$ is completely defined by the form
of the potential  $V(\theta; l, g)$ as a function of  $\theta$. Analysis of the
potential  $V(\theta; l, g)$ shows that its form essentially depends on values
of the parameters  $l$ and $g$. For $g < g_c$, the function $V(\theta; l, g)$,
like  $V_L(\theta; l)$ in the linear case, has one minimum for a certain
$\theta_1^{st}$. With increase in  $l$, the value $\theta_1^{st}$ increases and
the minimum of  $V(\theta_1^{st}; l, g)$ decreases. The behavior of the
potential  $V(\theta; l, g)$ essentially changes for  $g > g_c$, $l^b_1 < l <
l^b_2$. In this case, the function $V(\theta; l, g)$ has the form of a double
well with local minima at  $\theta = \theta_1^{st}$ and $\theta =
\theta_3^{st}$ separated by a maximum at $\theta = \theta_2^{st}$. As the
concentration $l$ varies from $l^b_1$ to $l^b_2$, the positions of the extrema,
the depths of the wells, and the barrier between them $\delta V_{2,1}(l, g) =
V(\theta_2^{st}; l, g) - V(\theta_1^{st}; l, g)$ vary. For concentrations near
$l^b_1$, the second well with minimum at  $\theta = \theta_3^{st}$ is rather
flat, essentially shallower as compared with the first well with minimum at
$\theta = \theta_1^{st}$, $V(\theta_3^{st}; l, g) > V(\theta_1^{st}, l, g)$,
and corresponds to a possible metastable state of the system. As the
concentration increases, the second well becomes deeper and the barrier between
the wells decreases. At a certain concentration, the depth of the second well
becomes equal to the depth of the first one. For higher concentrations, the
second well is deeper than the first, $V(\theta_3^{st}; l, g) <
V(\theta_1^{st}; l, g)$, i.e., the system is in a metastable state in the first
well and in a stable state in the second well. With a further increase in the
concentration, the modulus of the difference between the minima of the wells
$|V(\theta_3^{st}; l, g) - V(\theta_1^{st}; l, g)|$ increases and the barrier
$\delta V_{2,1}(l, g)$ decreases (the slope of the first well between
$\theta_1^{st}$ and  $\theta_2^{st}$ is close to zero) and disappears for $ l =
l^b_2$. As soon as the concentration becomes greater than the bifurcation value
$l^b_2$, the first minimum disappears, furthermore, for concentrations near
$l^b_2$ ($l
> l^b_2$), the potential $V(\theta; l, g)$, in a neighborhood of  $\theta_2^b$,
has almost a zero slope, which leads to the well-known critical slowing down of
the system \cite{ref.Gil,ref.Hak,ref.Sug}. The behavior of this gradient
dynamical system corresponds to the well-known principle of perfect delay
\cite{ref.Gil} according to which a transition of the bistable system between
two stable states of equilibrium is absent.

\subsection{Influence of the Masses of Center and Molecule on the Kinetics
of the Surface Coverage }  \label{Mass}

Now we investigate the kinetics of the surface coverage taking into account the
masses of adsorption center and molecule. The estimating condition (19) for the
overdamped approximation can be represented in the form $Q^2_M \ll 1$, where
$Q_M = \omega_M/2\, \gamma_M = \tau_M/\tau_r$ and $\gamma_M = \alpha/2M$ are,
respectively, the $Q$-factor and the damping constant of a free oscillator of
mass $M$. We also introduce the dimensionless quantities $\mu = m/m_0$  and $d
= (\tau_0/\tau_d)^2$, where $\tau_0 = 1/\omega_0$ and $\omega_0 =
\sqrt{\kappa/m_0}$ is the oscillation frequency of a free oscillator of mass
$m_0$. The quantity $Q_M$ can be represented in the form $Q_M = Q_0 \, \sqrt{1
+ \mu}$, where $Q_0 = \omega_0/2\, \gamma_0 = \tau_0/\tau_r = \sqrt{d}/\beta$
and $\gamma_0 = \alpha/2m_0$  are, respectively, the $Q$-factor and the damping
constant of a free oscillator of mass $m_0$.

Analysis of the singular points  $\theta^{st}_1$, $\theta^{st}_2$, and
$\theta^{st}_3$ of the system of equations (10)--(11) carried out in the phase
space gives the following:  For very low values of the $Q$-factor $Q_M$,
$\theta^{st}_1$, $\theta^{st}_3$ and $\theta^{st}_2$ are, as in the overdamped
case, stable nodes and a saddle, respectively \cite{ref.Kak}. With increase in
the $Q$-factor $Q_M$, starting from certain values that depend on values of the
parameters $a_g, \l, \beta$, and $\mu$, $\theta^{st}_1$ and $\theta^{st}_3$
become stable node-focuses. In the interval $(\theta^b_2, \, \theta^b_1)$, near
$\theta^b_1$, there appears a domain such that if $\theta^{st}_2$ falls within
this domain, then the singular point $\theta^{st}_2$ is a saddle-focus,
otherwise, it is a saddle. With increase in the $Q$-factor, this domain rapidly
grows and covers the entire interval $(\theta^b_2, \, \theta^b_1)$ so that the
unstable singular point $\theta^{st}_2$ is always a saddle-focus.

Below, we give the results of numerical analysis obtained for  $a_g = 1.1$,
i.e., for a system with possible bistability, and  $\beta = 1$.

\begin{figure}
 \centering{
 \epsfig{figure=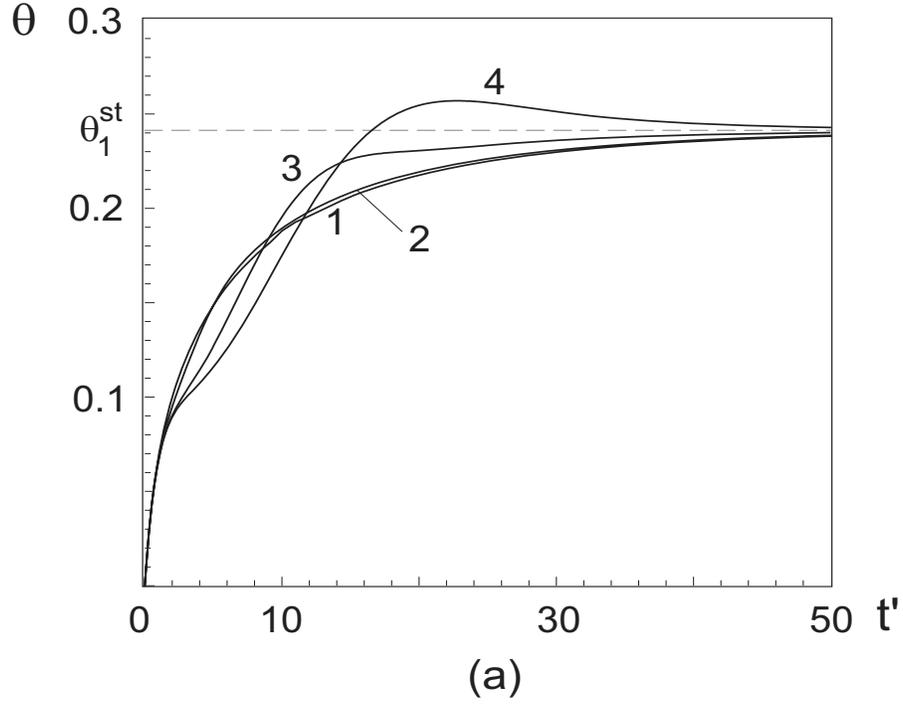, width=12cm, height=9.5cm} \\
 \vspace{1cm}
 \epsfig{figure=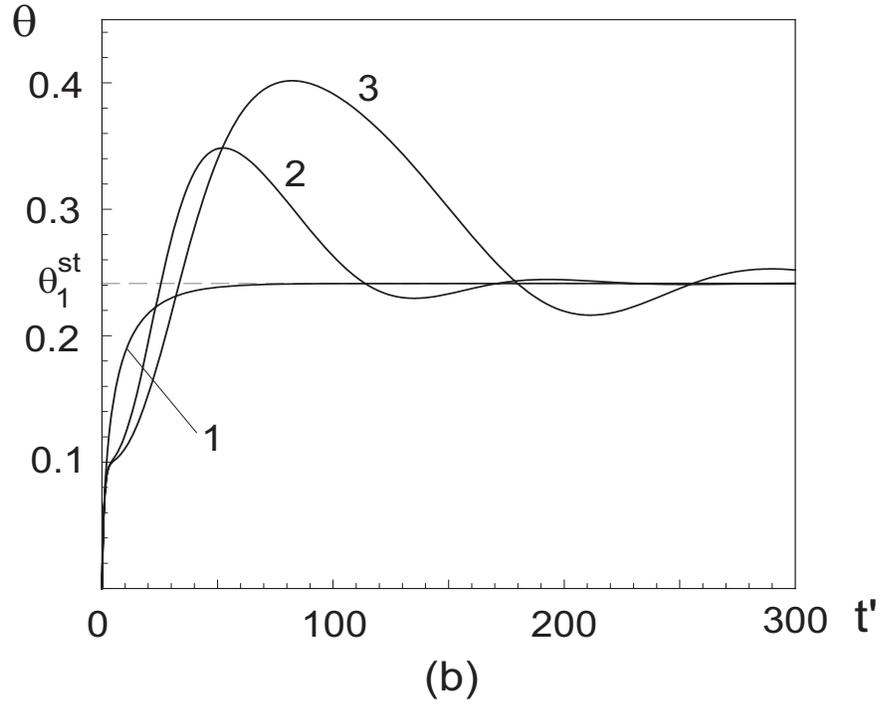, width=12cm, height=9.5cm} \\
 \bigskip
 \caption {Influence of the masses of adsorption center and molecule on the kinetics of
  the surface coverage  $\theta(t')$  for the concentration $l = 0.11$ lying in
  the middle of the bistability interval;
  $a_g = 1.1$, $\beta = 1$; $m_0 = 0, m = 0$ (curve 1),
  $\mu = 1$; (a) $d = $ 1 (2), 5 (3), 10 (4); (b) $d = $ 50 (2), 100 (3). }
 }
 \label{fig._7}
\end{figure}

The curves shown in Fig.~7 illustrate the influence of the masses of adsorption
center and molecule on the kinetics of the surface coverage  $\theta(t')$ for
the concentration  $l = 0.11$  lying in the middle of the bistability interval
$[l^b_1, \, l^b_2]$, where $l^b_1 \approx 0.1064$ and $l^b_2 \approx 0.1154$.
Here, for comparison, we present the quantity $\theta(t')$ without regard for
the masses of adsorption center and molecule (curve 1) and the stationary level
$\theta^{st}_1$ for this concentration (dashed line). For low values of masses
(curve 2 in Fig.~7a), as in the overdamped case, the number of molecules
adsorbed on the surface monotonically increases with time and reaches its
maximum value $\theta^{st}_1$. With increase in the coefficient $d$, which is
proportional to the mass of adsorption center, the behavior of the quantity
$\theta(t')$ changes. The surface coverage reaches its stationary value
$\theta^{st}_1$ only after several oscillations about it (Fig.~7b). The
amplitude and the number of oscillations as well as the time taken for
attaining the stationary value $\theta^{st}_1$ increase with $d$ (curves 2 and
3 in Fig.~7b). This behavior is caused by the inertia of the oscillator that
overshoots its equilibrium position and the deviation of the oscillator from
the equilibrium position increases with its mass. Therefore, taking account of
the masses of adsorption center and molecule changes only the character of
attainment of the nearest equilibrium state by the system.

\begin{figure}
 \centering{
 \epsfig{figure=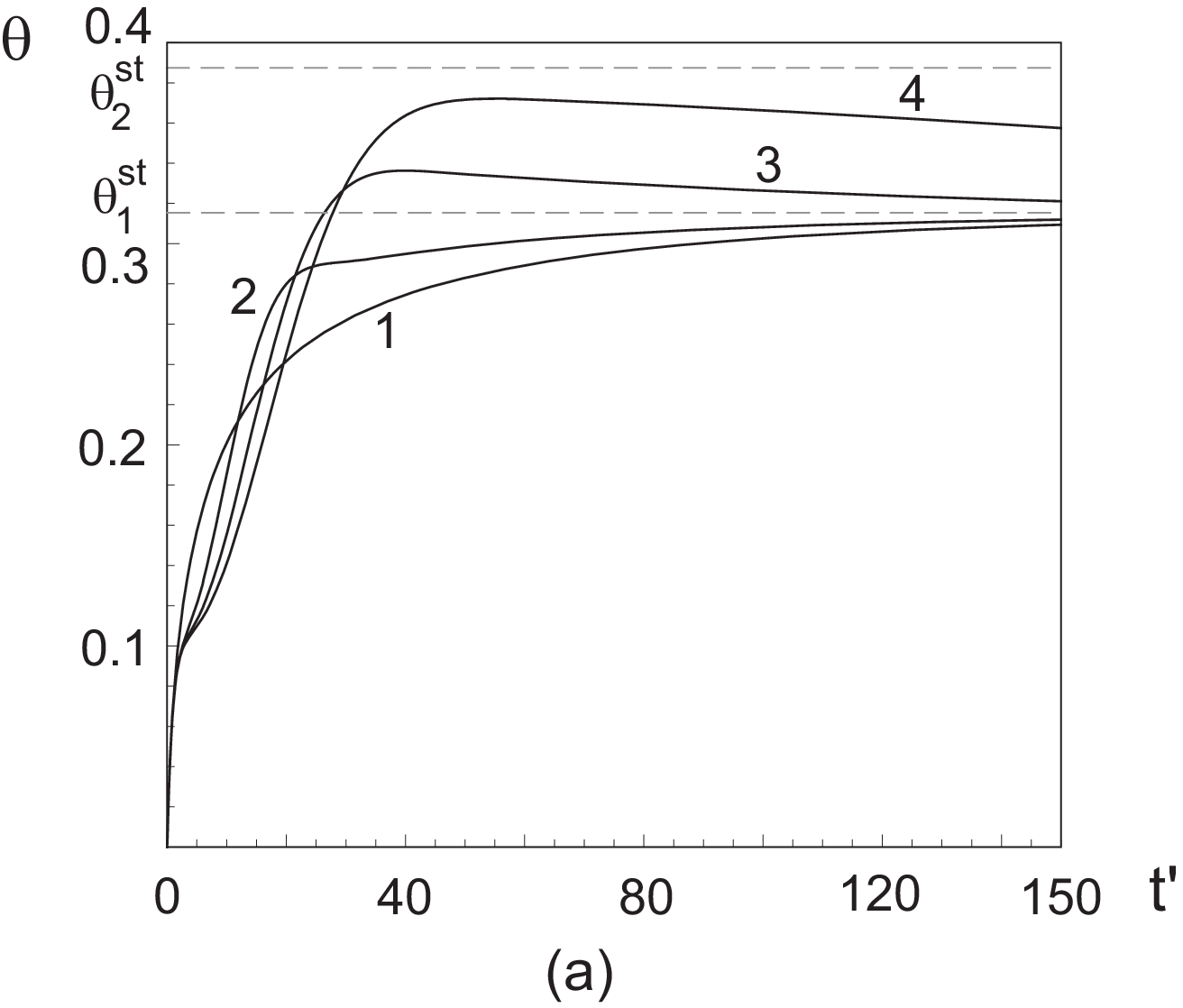, width=7.5cm, height=6.5cm}  \hspace{0.5cm}
 \epsfig{figure=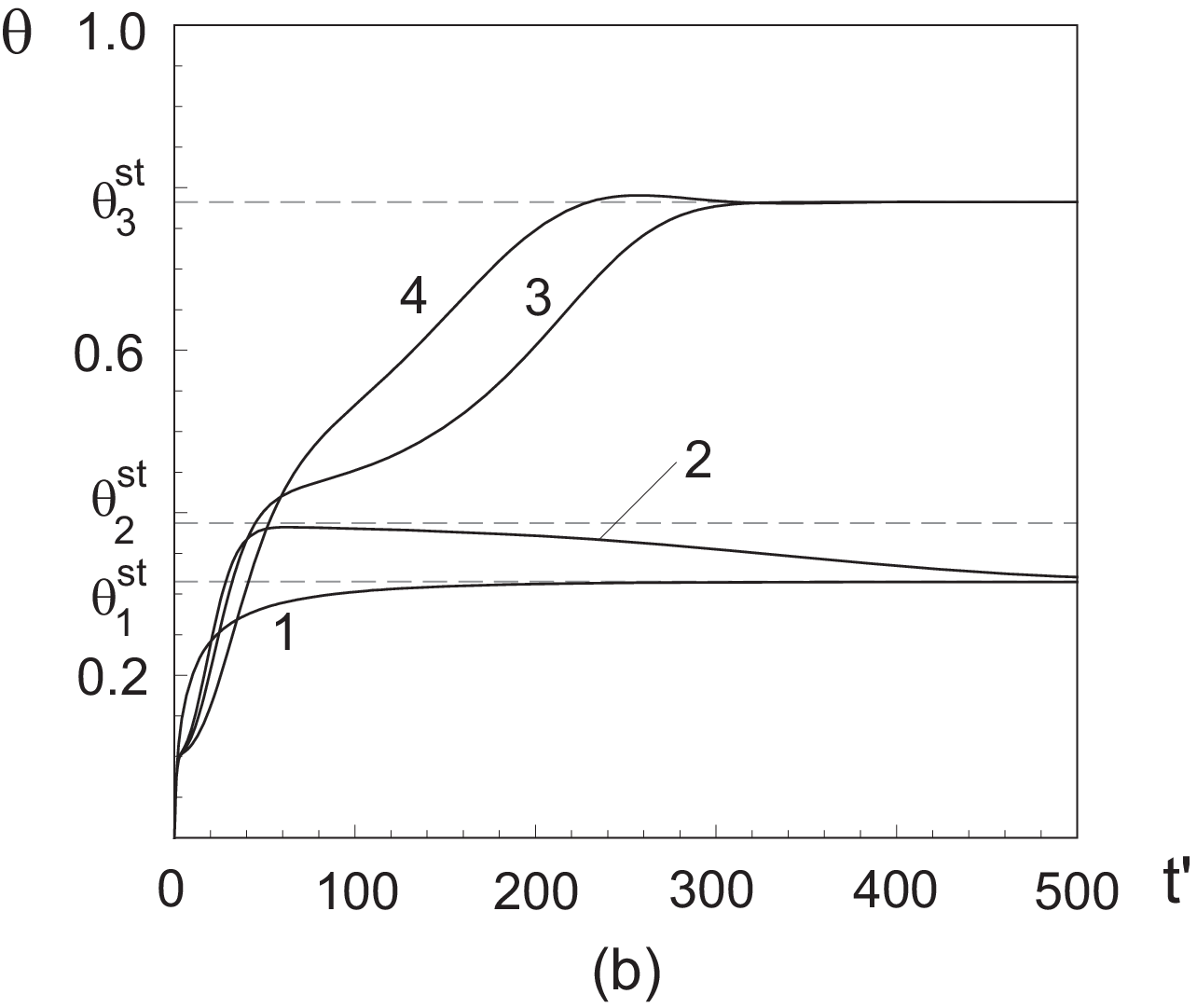, width=7.5cm, height=6.5cm} \\
  \vspace{0.6cm}
 \epsfig{figure=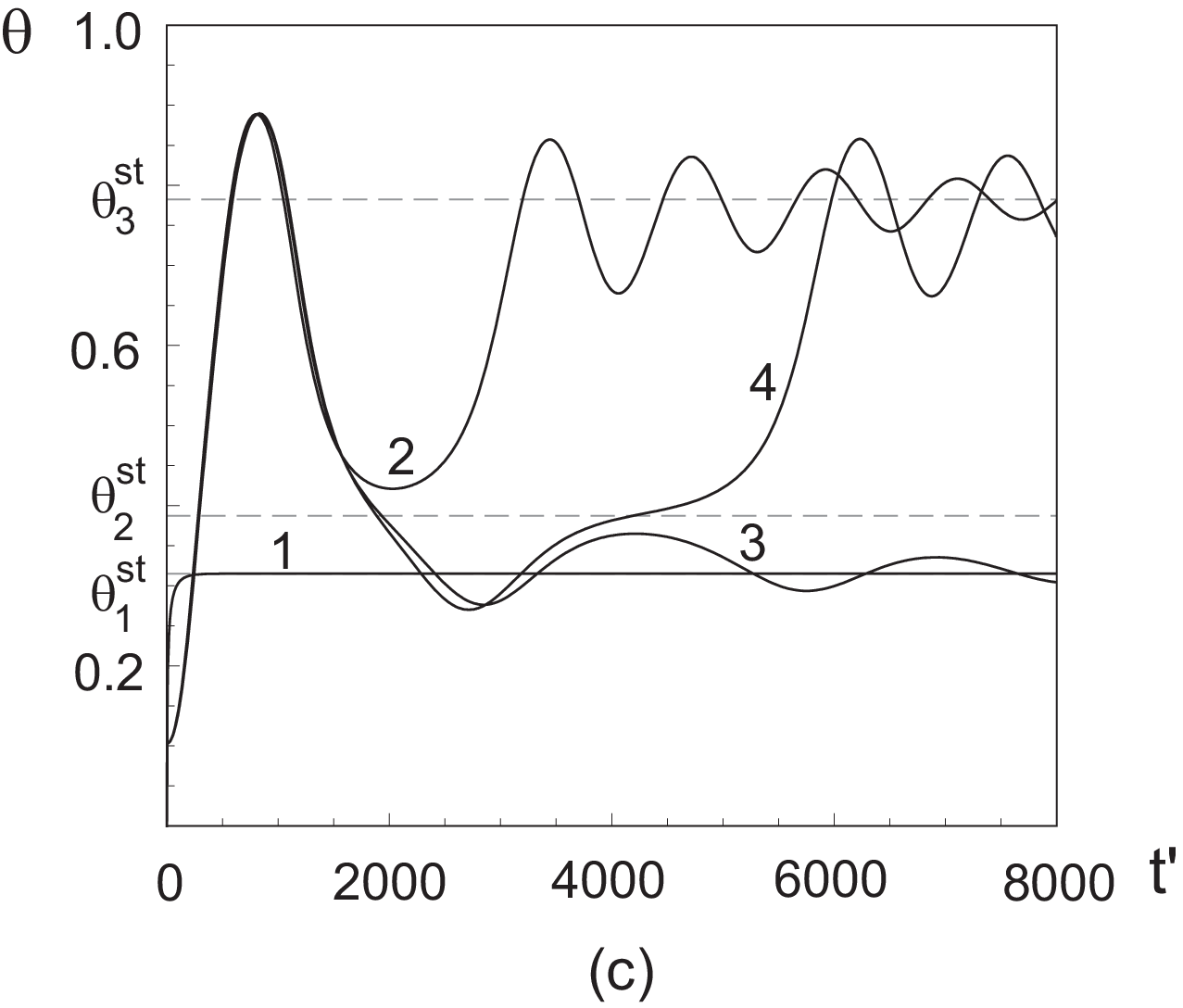, width=7.5cm, height=6.5cm} \\
  \bigskip
 \caption {Influence of the masses of adsorption center and molecule on the kinetics of
  the surface coverage $\theta(t')$  for the concentration $l = 0.115$ lying near
  the end point of the bistability interval;
  $a_g = 1.1$, $\beta = 1$; $m_0 = 0, m = 0$  (curve 1), $\mu = 1$;
  (a) $d = $ 10 (2), 20 (3), 30 (4);
  (b) $d = $ 33 (2), 50 (3), 100 (4);
  (c) $d = $ 4900 (2), 5200 (3), 5300 (4). }
 }
 \label{fig._8}
\end{figure}

The behavior of the system can essentially change if the concentration is near
the end point of the bistability interval. This case is shown in Fig.~8 where
the stationary states of the system (the stable states $\theta^{st}_1$ and
$\theta^{st}_3$ and the unstable state  $\theta^{st}_2$) are shown by dashed
lines. The curves  $\theta(t')$ in Fig.~8a show that, for low values of $d$,
the behavior of the system, to a large extent, is analogous to its behavior for
low masses considered above (Fig.~7a). Due to inertia, the system penetrates
into the domain $(\theta^{st}_1, \, \theta^{st}_2)$, which is a domain of
attraction of the attractor $\theta^{st}_1$ (domain I) \cite{ref.BaL}. As in
the overdamped case, with time, the system attains its stationary level
$\theta^{st}_1$. However, as soon as the mass of the complex reaches a value
for which the kinetic energy of the complex is sufficient to overcome the
``barrier'' $\theta^{st}_2$, the behavior of the system qualitatively changes
(cf. curves 2 and 3 in Fig.~8b). Having fallen into the domain $(\theta^{st}_2,
\, \theta^{st}_3)$, which is a domain of attraction of the attractor
$\theta^{st}_3$ (domain II), the system moves toward its second stable
stationary state  $\theta^{st}_3$. Having reached this state, the system
oscillates about it with decreasing amplitude. As $d$  increases, the time of
attainment of the stationary state  $\theta^{st}_3$  decreases because the
system, in fact, is not delayed in the neighborhood of the unstable state
$\theta^{st}_2$ (curve 4 in Fig.~8b). Thus, unlike the overdamped case for
which the stable equilibrium state $\theta^{st}_3$ is not attainable, due to
the masses of adsorption center and molecule, the system can be in this state
rather than in the state $\theta^{st}_1$.

A further increase in  $d$ is accompanied by an increase in the amplitude of
oscillations of the system about the stationary level $\theta^{st}_3$ and the
number of oscillations (high-$Q$ system). However, the system does not leave
domain II (curve 2 in Fig.~8c). If the amplitude of oscillations exceeds
$(\theta^{st}_3 - \theta^{st}_2)$, then the system falls into domain I and its
subsequent behavior can be different. The system can remain in this domain and,
after a time, it attains the stable stationary level  $\theta^{st}_1$ (curve 3
in Fig.~8c) as in the overdamped case (curve 1) and in the case of low values
of masses (curves 2--4 in Fig.~8a and curve 2 in Fig.~8b).

\begin{figure}
 \centering{
 \epsfig{figure=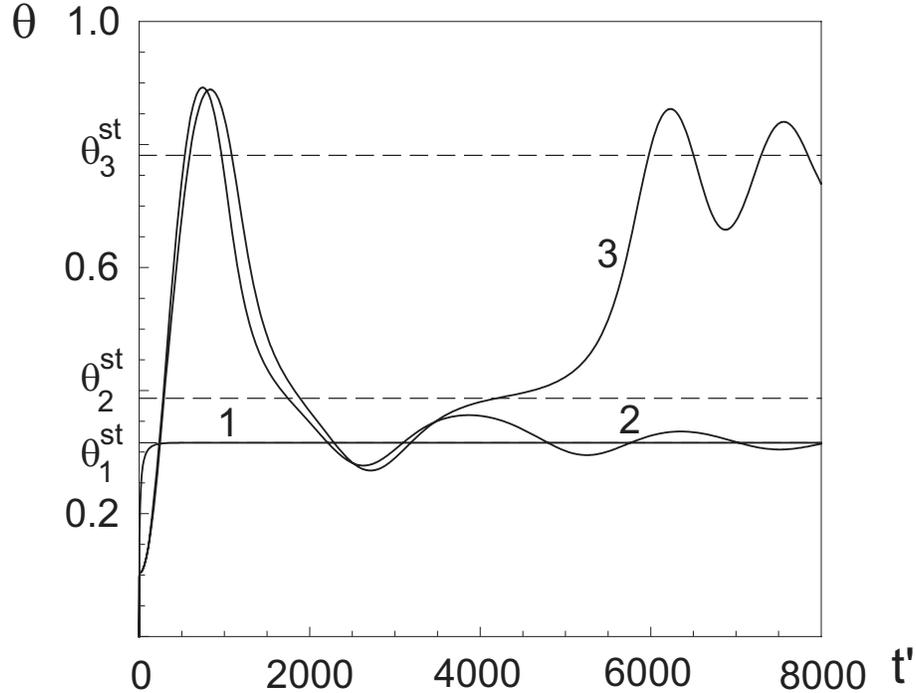, width=12cm, height=9.5cm} \\
 \bigskip
 \caption {Influence of the ratio of the masses of molecule and adsorption center on
  the kinetics of the surface coverage  $\theta(t')$;  $a_g = 1.1$, $\beta = 1$, $l = 0.115$;
  $m_0 = 0, m = 0$  (curve 1); $d = $ 5300, $\mu = $ 0.5 (2), 1 (3). }
 }
 \label{fig._9}
\end{figure}

\begin{figure}
 \centering{
 \epsfig{figure=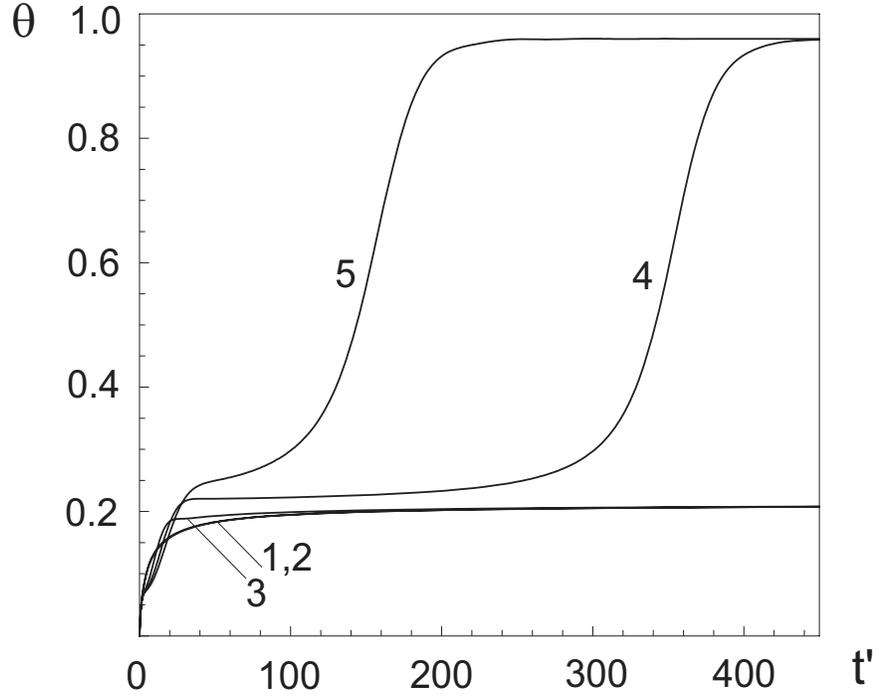, width=12cm, height=9.5cm} \\
 \vspace{1cm}
 \epsfig{figure=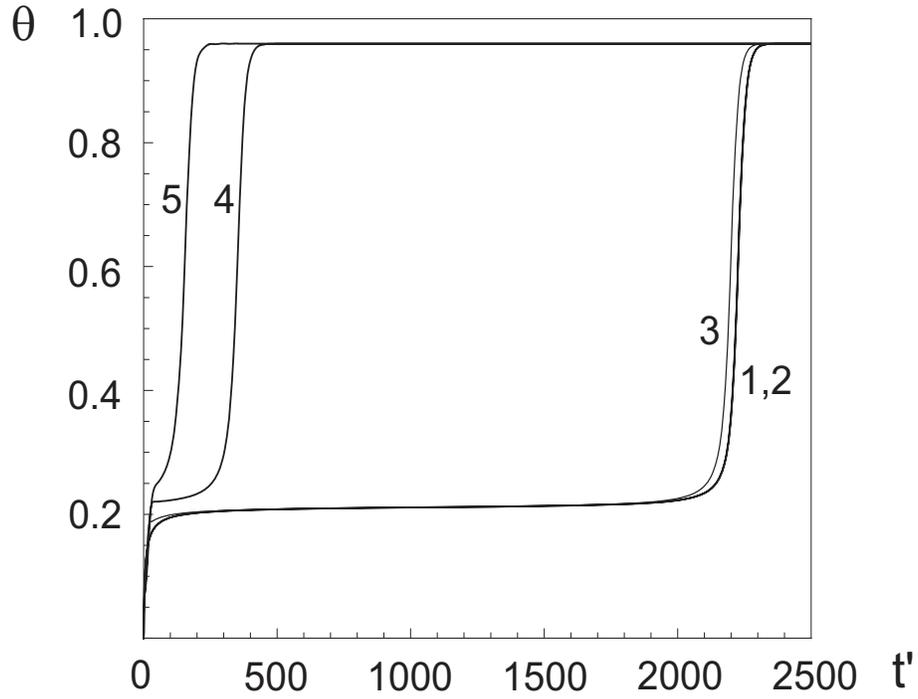, width=12cm, height=9.5cm} \\
 \bigskip
 \caption {Influence of the masses of adsorption center and molecule on the kinetics of the
  surface coverage  $\theta(t')$
  for the concentration close to the bifurcation concentration $l^b_2$;
  $a_g = 1.5$, $\beta = 1$, $\delta = 10^{-4}$; $m_0 = 0, m = 0$ (curve 1);
  $\mu = 1$, $d = $ 1 (2), 10 (3), 20 (4), 30 (5). }
 }
 \label{fig._10}
\end{figure}

For a somewhat greater value of  $d$, the kinetic energy of the complex can be
sufficient for the system to overcome the ``barrier'' $\theta^{st}_2$ for the
second time. As a result, the system again penetrates into domain II and
attains the stationary level $\theta^{st}_3$ (curve 4 in Fig.~8c). It is clear
that, with increase in mass (the value of $d$), the system can again return
into domain I, etc. Therefore, the finial stable state of the bistable system
($\theta^{st}_1$ or $\theta^{st}_3$) depends on the number of intersections of
the unstable state $\theta^{st}_2$ by the system in the process of its
evolution with time. The value of the parameter $\beta$, which is proportional
to the friction coefficient, also affects overcoming of the ``barrier''
$\theta^{st}_2$ by the system. Decreasing this parameter, i.e., increasing the
$Q$-factor of the system, it is possible, in principle, to realize a mode in
which the system visits each of two stable stationary states many times. This
behavior of the system under study qualitatively agrees with the well-known
behavior of a Newton gradient system whose potential energy has two minima
separated by a barrier in the case of a low value of the damping parameter
\cite{ref.Gil}.

Since the $Q$-factor of the system is determined both by the mass of adsorption
center and by the mass of adsorbed molecule, these two characteristics affect
(but different in rights) the possibility of the system to overcome the
``barrier'' $\theta^{st}_2$. The curves  $\theta(t')$ depicted in Fig.~9, which
describe the kinetics of the surface coverage for systems with equal masses of
adsorption centers but different masses of adsorbed molecules, visually
illustrate this conclusion. The behaviors of systems that returned from domain
II into domain I are different. The kinetic energy of the complex with more
light-weight molecule (curve 2) is insufficient for the complex to overcome the
``barrier'' $\theta^{st}_2$ for the second time, and the system is stabilized
at the stable level $\theta^{st}_1$. At the same time, the complex with heavier
molecule is able to overcome this ``barrier'' and the system returns into
domain II and attains the stable level $\theta^{st}_3$ (curve 3).

Note that the behavior of the kinetic curves shown in Figs.~7--9, on the
qualitative level, agrees with conclusion on the kinetics of the surface
coverage made in Appendix~B with the use of the effective potential in the
special case where the relaxation time of the quantity   $\theta(t)$ and the
characteristic times of the quantity $\xi(t)$ are essentially different.

For concentrations $l = l^b_2 \, \left(1 + \delta \right)$ that slightly exceed
the bifurcation value $l^b_2$, taking account of the masses of adsorption
center and molecule can also qualitatively change the kinetics of the surface
coverage (Fig.~10). For equal masses of adsorption center and molecule ($\mu =
1$) and for values  $d \leq 1$ (i.e., for  $Q_0 \leq 1$), the influence of
masses on the behavior of $\theta(t')$ is insignificant (curves 1 and 2, in
fact, coincide). With increase in the mass, the plot of the function
$\theta(t')$ in the form of a double step typical of the overdamped case
remains true (curve 3 in Fig.~10b). However, the residence time of the system
in the ``quasistationary'' state somewhat decreases, i.e., the delay of the
system in a neighborhood of the singular point  $\theta^b_2$ corresponding to
the bifurcation concentration $l^b_2$ is less than in the overdamped case. As
the mass  increases, the residence time of the system in the
``quasistationary'' state rapidly decreases (curve 4 in Fig.~10b) and, starting
from certain values of  $d$,  the system moves to the stable stationary state
$\theta^{st}_1$ without delay near the ``quasistationary '' state $\theta^b_2$
(curve 5). Thus, the possibility of an intermediate ``quasistationary'' state
for the system essentially depends on relations between the inertial and
dissipative characteristics of the system.



\section{Conclusions }  \label{Conclusions}

In the present paper, we have proposed the self-consistent model taking into
account variations in adsorption properties of the adsorbent surface in the
process of adsorption--desorption of molecules of gas on it. Within the
framework of this model, we have introduced the dimensionless coupling
parameter that characterizes the interaction of an adsorbed molecule with
polarized medium. We have established that the system can be bistable if this
parameter is greater than critical and the concentration of molecules in the
gas phase belongs to a certain interval. We have investigated bifurcation
concentrations for which stable states of the system appear and disappear. We
have obtained adsorption isotherms that essentially differ from the classical
Langmuir isotherms. It is established the possibility of the Zeldovich
hysteresis within the framework of the proposed model. It is shown that
variations in adsorption properties of the surface in the course of
adsorption--desorption can lead to a peculiar adaptation of the system to the
state in which the majority of adsorption centers is bound up to very low
concentrations.

The detailed analysis of the kinetics of the surface coverage established that
taking account of variations in adsorption properties of the surface in the
course of adsorption--desorption leads to new phenomena: a ``quasistationary''
state for the overdamped approximation and damped self-oscillations of the
system in the general case.

\begin{acknowledgments}
The author expresses the deep gratitude to Prof. Yu.\,B.~Gaididei for the
statement of the problem, valuable remarks, and useful discussions of results.
\end{acknowledgments}


\appendix*

\section{A}  \label{Application}

A change in the activation energy for adsorption  $E_a$ caused by polarization
of the medium in the process of adsorption--desorption depends on various
factors connected both with adsorbent and adsorbate. As an example, we consider
the case where, due to polarization of the medium, the activation energy for
adsorption $E_a$ decreases by the quantity  $\chi \,x$ equal to the increment
of the activation energy for desorption  $E_d$ caused by polarization.
Supposing that the preexponential factor $k_+$ is not changed, we obtain the
following expression for the adsorption rate characteristic $k_a(x)$:

\begin{equation}
 k_a(x) = k_a \exp{\biggl(\frac{\chi \, x}{k_B T} \biggr)},
\end{equation}

\noindent which, like the desorption rate characteristic  $k_d(x)$ defined by
relation (9), depends on the concentration of gas. As a result, we obtain a
system of equations that describes the kinetics of the quantity of adsorbed
substance and differs from system (10)--(11) derived above under the assumption
that the activation energy for adsorption does not vary in the process of
adsorption--desorption only by the replacement of Eq.~(10) by the equation

\begin{equation}
 \frac{d \theta}{dt} = k_a C \bigl( 1 - \theta \bigr) \exp{\left(g\, \xi \right)}
  - k_d \, \theta \exp{\left(-g\, \xi \right)}.
\end{equation}

\noindent The factor $ \exp{\left(g\, \xi \right)}$ in the first term on the
right-hand side of this equation takes into account a change in the activation
energy for adsorption in the process of adsorption--desorption of molecules of
gas.

In the stationary case, we obtain the same equation (12) for determination of
the quantity  $\theta^{st}$ but with function  $F(\theta)$ with changed
parameter $g$

\begin{equation}
 F(\theta) = \frac{\theta}{1 - \theta} \, \exp{\Bigl( - 2g\, \theta \Bigr)}.
\end{equation}

In addition, we obtain

\begin{equation}
 k_a(\theta) = k_a \, \exp \bigl(g\, \theta \bigr),  \qquad \qquad
 K(\theta) = \frac{k_a(\theta)}{k_d(\theta)} = K \, \exp \bigl( 2g\, \theta
 \bigr).
\end{equation}

Comparing (A4) and (15), we obtain a natural result that a decrease in the
activation barrier in the process of adsorption--desorption leads to a shift of
the equilibrium of the system towards an increase in the number of adsorbed
molecules.

Since the behavior of function (A.3) is identical to the behavior of function
(12) with replacement of the critical value of the coupling parameter $g_c = 4$
by $g_c = 2$, the results of analysis of adsorption isotherms carried out in
the the third section of the present paper remains also true in the case at
hand with replacement $g_c \rightarrow g_c/2$. By analogy, the critical
temperature below which the system can be bistable is changed, $T_c \rightarrow
2 T_c$.

The kinetics of the surface coverage $\theta(t')$ is analogous to the kinetics
of  $\theta(t')$ established above without regard for a change in the
activation energy for adsorption (Figs.~4, 5, and 7--10). However, in the
considered case, it is somewhat faster, which is quite natural because a
decrease in the barrier  $E_a$ favors a faster filling of the surface with
molecules of gas. The specific features of the kinetics of $\theta(t')$
depicted in Figs.~8 and 9 for a bistable system also occur, furthermore, they
are realized for lesser values of masses of adsorption centers and molecules.


\section{B}  \label{Application B}

Here, we investigate the behavior of the dynamical system (10)--(11) that
describes the kinetics of adsorption of molecules on the surface whose
adsorption properties vary in the process of adsorption--desorption in the
special case where the relaxation time of the quantity $\theta(t)$ is much less
than the characteristic times of the quantity $\xi(t)$, i.e., the variables
$\theta$ and $\xi$ are fast and slow, respectively. Performing the adiabatic
elimination of the fast variable $\theta(t)$  \cite{ref.Hak}, namely, setting $
d\theta/dt = 0$ in Eqs.~(10) and (11), we obtain the following representation
for the surface coverage versus the slow variable $\xi$:

$$
 \theta = \frac{l}{l + \exp{\left(-g\, \xi\, \right)}}.   \eqno (B.1)
$$

The dimensionless coordinate of oscillator  $\xi(t)$ is determined as a
solution of the nonlinear differential equation

$$
 m_{eff}(\xi)\, \frac{d^2\xi}{dt^2} + \alpha \frac{d\xi}{dt} = - \frac{dU(\xi)}{d\xi}
     \eqno (B.2)
$$
that describes the motion of the oscillator with effective variable mass

$$
 m_{eff}(\xi) = m_L  + m \, \theta_L\, \frac{1 - \exp{\left(-g\, \xi\, \right)}}
                {l + \exp{\left(-g\, \xi\, \right)}},    \eqno (B.3)
$$

$$
 m_L = m_0 + m \, \theta_L,    \eqno (B.4)
$$
in the effective potential

$$
 U(\xi) = \frac{\kappa}{2} \, \biggl\{ \xi^2 - 2\,\xi - \frac{2}{g}
   \ln{\frac{l + \exp{\left(-g\, \xi\, \right)}}{l + 1}} \biggr\}.    \eqno (B.5)
$$
Note that the second term in relation (B.3) for the effective mass disappears
in the absence of polarization of the adsorbate in the process of
adsorption--desorption, i.e.,  $\lim\limits_{g \rightarrow 0}\, m_{eff}(\xi) =
m_L$.

Therefore, we reduced the problem of investigation of the kinetics of the
surface coverage to the problem of study of the motion of an oscillator of
variable mass in potential (B.5). Since the quantity $\xi$ is the dimensional
coordinate of a bound adsorption center, in terms of the coordinate $x$ of this
center, the equation of motion for it has the form

$$
 m_{eff}(x)\, \frac{d^2 x}{dt^2} + \alpha \frac{dx}{dt} = - \frac{d U(x)}{dx},
     \eqno (B.6)
$$
where

$$
 m_{eff}(x) = m_L  + m \, \theta_L\, \frac{1 - \exp{\left(-b\, x\, \right)}}
                {l + \exp{\left(-b\, x\, \right)}},    \eqno (B.7)
$$

$$
 U(x) = \frac{\kappa\, x^2}{2} -  \chi \, x - k_B T
        \ln{\frac{l + \exp{\left(-b\, x\, \right)}}{l + 1}} \biggr\},    \eqno (B.8)
$$

$$
 b = \frac{g}{x_{max}} = \frac{\chi}{k_B T}.    \eqno (B.9)
$$

Note that the effective potential (B.8) is analogous to the potential derived
in the adiabatic approximation in  \cite{ref.Chr1} where the structural
regulation of functioning of a macromolecular in repeating cycles of reactions
is investigated.

Analysis of the potential  $U(\xi)$ shows that, for  $g> g_c$,  $l^b_1 < l <
l^b_2$, it has the form of a double well with local minima  at  $\xi =
\xi_1^{st}$ and $\xi = \xi_3^{st}$ separated by a maximum at $\xi =\xi_2^{st}$,
where $\xi_n^{st} = \theta_n^{st}$, $n = 1,2,3$, and $\theta_n^{st}$ are the
stationary surface coverages investigated in Sec.~3 that satisfy Eq.~(12). For
$g < g_c$ and any $l$  as well as for  $g> g_c$ and $l < l^b_1$ or $l > l^b_2$,
the potential $U(\xi)$ has one minimum.

\begin{figure}
 \centering{
 \includegraphics[width=0.9\textwidth]{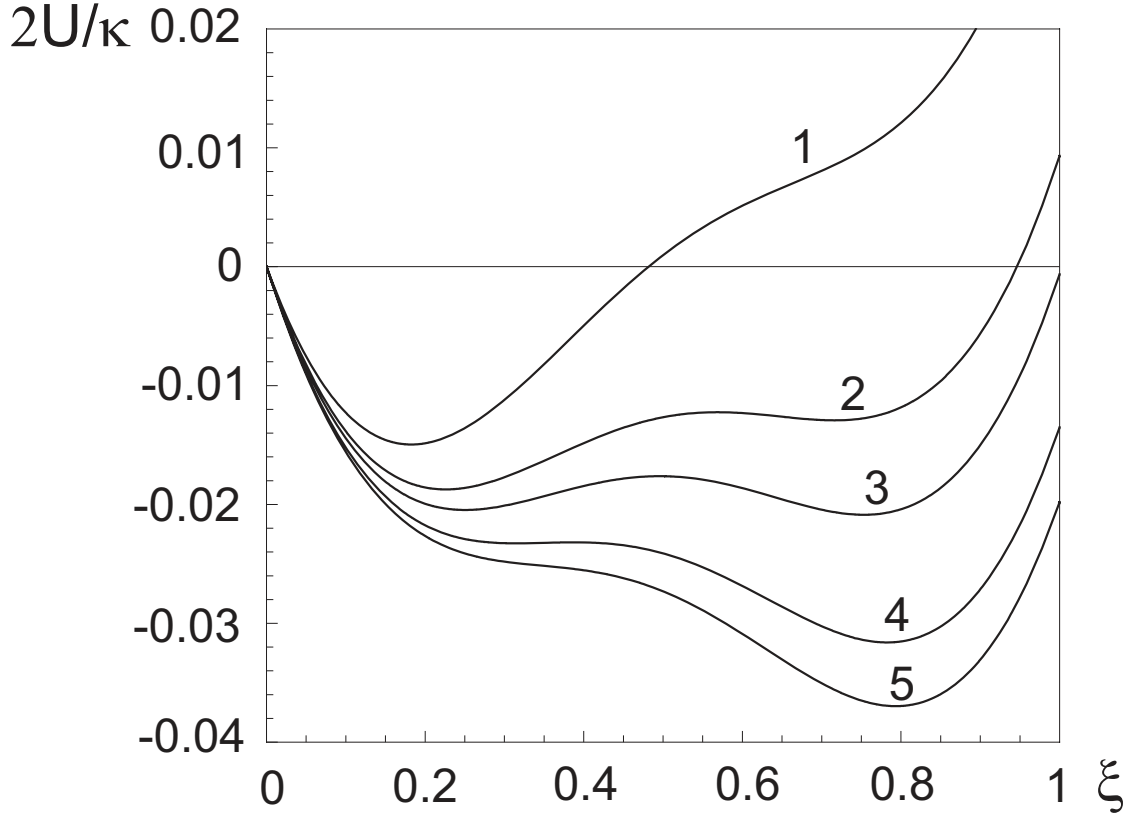}
 \bigskip
 \caption {Normed effective potential for different concentrations:  $g = 4.4$,
  $l = $ 0.1 (1), 0.108 (2), 0.11 (3), 0.115 (4), 0.117 (5). }
 }
 \label{fig._11}
\end{figure}

The curves presented in Fig.~11 for $g = 4.4 > g_c$ clearly illustrate the
essential influence of the concentration on the form of the potential. For
concentrations lying outside the interval  $[l^b_1, \, l^b_2]$, the potential
has a single minimum (curves 1 (for $l < l^b_1 \approx 0.1064$) and 5 (for $l >
l^b_2 \approx 0.1154$)), furthermore, the equilibrium position of the
oscillator for $l > l^b_2$ is considerably more distant from the nonperturbed
position $\xi = 0$ than that for  $l < l^b_1$. Curves 2--4 illustrate the
double-well character of the potential for the concentrations $l^b_1 < l <
l^b_2$ and deepening of its wells (especially, the second well) with increase
in the concentration.

In the case of the double-well potential $U(\xi)$, the motion of the oscillator
described by Eq.~(B.2), which was initially at rest at the point $\xi = 0$, can
be different depending on the contributions of the inertial and dissipative
terms. For small masses and large values of the friction coefficient, the
oscillator rolls down into the nearest well of the potential $U(\xi)$ and,
finally, is stabilized in the steady state at the point $\xi_1^{st}$
corresponding to a minimum of the potential. With increase in mass and/or a
decrease in the friction coefficient, the kinetic energy of the oscillator may
be sufficient to overcome the potential barrier between the wells and the
oscillator falls in the second well. Depending on the values of masses and the
friction coefficient, the oscillator can both remain in the second well with
subsequent stabilization at its minimum and return to the first well. For very
small values of $\alpha$, the oscillator can many times visit each well before
stabilization in one of them.

The surface coverage  $\theta(t)$ has a similar behavior. Therefore, the
kinetics of the surface coverage shown in Figs.~7--9 and obtained without
additional assumptions on fast and slow variables, on the qualitative level,
agrees with conclusions made above on the basis of the motion of an oscillator
in a double-well potential.



\end{document}